\documentclass[twocolumn]{aastex631}

\usepackage{newtxtext,newtxmath,comment} 
\usepackage[normalem]{ulem}

\usepackage[T1]{fontenc}
\usepackage{ae,aecompl}
\usepackage{xcolor}
\usepackage{graphicx}
\usepackage{amsmath}
\usepackage{enumitem}
\usepackage{url}
\usepackage{tasks}

\def\name{the DREAMS project }
\def\hp{h_\mathrm{peak}}
\def\kp{k_\mathrm{peak}}
\def\hpkp{\hp-\kp}

\makeatletter
\let\frontmatter@title@above=\relax
\makeatother

\begin{document}

\title[The DREAMS Project]{Introducing the DREAMS Project: \\
DaRk mattEr and Astrophysics with Machine learning and Simulations} 

\correspondingauthor{Jonah C. Rose} \\
\email{j.rose@ufl.edu}

\author{Jonah C. Rose}
\affiliation{Department of Astronomy, University of Florida, Gainesville, FL 32611, USA}
\affiliation{Center for Computational Astrophysics, Flatiron Institute, 162 5th Avenue, New York, NY 10010, USA}

\author{Paul Torrey}
\affiliation{Department of Astronomy, University of Virginia, Charlottesville, VA 22904, USA}

\author{Francisco Villaescusa-Navarro}
\affiliation{Center for Computational Astrophysics, Flatiron Institute, 162 5th Avenue, New York, NY 10010, USA}
\affiliation{Department of Astrophysical Sciences, Princeton University, Peyton Hall, Princeton, NJ 08544, USA}

\author{Mariangela Lisanti}
\affiliation{Center for Computational Astrophysics, Flatiron Institute, 162 5th Avenue, New York, NY 10010, USA}
\affiliation{Department of Physics, Princeton University, Princeton, NJ 08544, USA}

\author{Tri Nguyen}
\affiliation{Department of Physics and Kavli Institute for Astrophysics and Space Research, Massachusetts Institute of Technology, Cambridge, MA 02139, USA}
\affiliation{The NSF AI Institute for Artificial Intelligence and Fundamental Interactions, Cambridge, MA 02139, USA}

\author{Sandip Roy}
\affiliation{Department of Physics, Princeton University, Princeton, NJ 08544, USA}

\author{Kassidy E. Kollmann}
\affiliation{Department of Physics, Princeton University, Princeton, NJ 08544, USA}

\author{Mark Vogelsberger}
\affiliation{Department of Physics and Kavli Institute for Astrophysics and Space Research, Massachusetts Institute of Technology, Cambridge, MA 02139, USA}
\affiliation{The NSF AI Institute for Artificial Intelligence and Fundamental Interactions, Cambridge, MA 02139, USA}

\author{Francis-Yan Cyr-Racine}
\affiliation{Department of Physics and Astronomy, University of New Mexico, 210 Yale Blvd NE, Albuquerque, NM 87106, USA}

\author{Mikhail V. Medvedev}
\affiliation{Department of Astrophysical Sciences, Princeton University, Peyton Hall, Princeton, NJ 08544, USA}
\affiliation{Institute for Advanced Study, School for Natural Sciences, Princeton, NJ 08540}
\affiliation{Institute for Theory and Computation, Harvard University, Cambridge, MA 02138}
\affiliation{Department of Physics and Astronomy, University of Kansas, Lawrence, KS 66045}
\affiliation{Laboratory for Nuclear Science, Massachusetts Institute of Technology, Cambridge, MA 02139}

\author{Shy Genel}
\affiliation{Center for Computational Astrophysics, Flatiron Institute, 162 5th Avenue, New York, NY 10010, USA}
\affiliation{Columbia Astrophysics Laboratory, Columbia University, 550 West 120th Street, New York, NY 10027, USA}

\author{Daniel Angl\'es-Alc\'azar}
\affiliation{Department of Physics, University of Connecticut, 196 Auditorium Road, U-3046, Storrs, CT 06269-3046, USA}
\affiliation{Center for Computational Astrophysics, Flatiron Institute, 162 5th Avenue, New York, NY 10010, USA}

\author{Nitya Kallivayalil}
\affiliation{Department of Astronomy, University of Virginia, Charlottesville, VA 22904, USA}

\author{Bonny Y. Wang}
\affiliation{McWilliams Center for Cosmology and Astrophysics, Carnegie Mellon University, Pittsburgh, PA 15213}

\author{Bel\'en Costanza}
\affiliation{Facultad de Ciencias Astron\'omicas y Geof\'isicas, Universidad Nacional de La Plata, Observatorio Astron\'omico, Paseo del Bosque,  B1900FWA La Plata, Argentina}

\author{Stephanie O'Neil}
\affiliation{Department of Physics and Kavli Institute for Astrophysics and Space Research, Massachusetts Institute of Technology, Cambridge, MA 02139, USA}

\author{Cian Roche}
\affiliation{Department of Physics and Kavli Institute for Astrophysics and Space Research, Massachusetts Institute of Technology, Cambridge, MA 02139, USA}

\author{Soumyodipta Karmakar}
\affiliation{Department of Physics and Astronomy, University of New Mexico, 210 Yale Blvd NE, Albuquerque, NM 87106, USA}

\author{Alex M. Garcia}
\affiliation{Department of Astronomy, University of Virginia, Charlottesville, VA 22904, USA}

\author{Ryan Low}
\affiliation{Department of Physics and Astronomy, University of Kansas, Lawrence, KS 66045}

\author{Shurui Lin}
\affiliation{Department of Astronomy, School of Physical Sciences, University of Science and Technology of China, Hefei, Anhui 230026, China}
\affiliation{CAS Key Laboratory for Researches in Galaxies and Cosmology, School of Astronomy and Space Science, University of Science and Technology of China, Hefei, Anhui 230026, China}

\author{Olivia Mostow}
\affiliation{Department of Astronomy, University of Virginia, Charlottesville, VA 22904, USA}

\author{Akaxia Cruz}
\affiliation{Center for Computational Astrophysics, Flatiron Institute, 162 5th Avenue, New York, NY 10010, USA}
\affiliation{Department of Physics, Princeton University, Princeton, NJ 08544, USA}

\author{Andrea Caputo}
\affiliation{Theoretical Physics Department, CERN, 1211 Geneva 23, Switzerland}

\author{Arya~Farahi}
\affiliation{Departments of Statistics and Data Science, University of Texas at Austin, Austin, TX 78757, USA}

\author{Julian B.~Mu\~noz}
\affiliation{Department of Astronomy, University of Texas at Austin, Austin, TX 78757, USA}

\author{Lina Necib}
\affiliation{Department of Physics and Kavli Institute for Astrophysics and Space Research, Massachusetts Institute of Technology, Cambridge, MA 02139, USA}
\affiliation{The NSF AI Institute for Artificial Intelligence and Fundamental Interactions, Cambridge, MA 02139, USA}

\author{Romain Teyssier}
\affiliation{Department of Astrophysical Sciences, Princeton University, Peyton Hall, Princeton, NJ 08544, USA}

\author{Julianne J.~Dalcanton}
\affiliation{Center for Computational Astrophysics, Flatiron Institute, 162 5th Avenue, New York, NY 10010, USA}
\affiliation{Department of Astronomy, University of Washington, Box 351580, Seattle, WA 98195, USA}

\author{David Spergel}
\affiliation{Center for Computational Astrophysics, Flatiron Institute, 162 5th Avenue, New York, NY 10010, USA}

\begin{abstract}
We introduce the DREAMS project, an innovative approach to understanding the astrophysical implications of alternative dark matter models and their effects on galaxy formation and evolution.
The DREAMS project will ultimately comprise thousands of cosmological hydrodynamic simulations that simultaneously vary over dark matter physics, astrophysics, and cosmology in modeling a range of systems---from galaxy clusters to ultra-faint satellites.  Such extensive simulation suites can provide adequate training sets for machine-learning-based analyses.  
This paper introduces two new cosmological hydrodynamical suites of Warm Dark Matter, each comprised of 1024 simulations generated using the \textsc{arepo} code.  One suite consists of uniform-box simulations covering a $(25~h^{-1}~{\rm M}_\odot)^3$ volume, while the other consists of Milky Way zoom-ins with sufficient resolution to capture the properties of classical satellites. 
 For each simulation, the Warm Dark Matter particle mass is varied along with the initial density field and several parameters controlling the strength of baryonic feedback within the IllustrisTNG model. We provide two examples, separately utilizing emulators and Convolutional Neural Networks, to demonstrate how such simulation suites can be used to disentangle the effects of dark matter and baryonic physics on galactic properties. The DREAMS project can be extended further to include different dark matter models, galaxy formation physics, and astrophysical targets.  In this way, it will provide an unparalleled opportunity to characterize uncertainties on predictions for small-scale observables, leading to robust predictions for testing the particle physics nature of dark matter on these scales. 
\end{abstract}


\section{Introduction}
\label{sec:intro}

The DaRk mattEr and Astrophysics with Machine learning and Simulations (DREAMS\footnote{\url{https://dreams-project.readthedocs.io}\\ \url{https://www.dreams-project.org}}) project aims to disentangle how variations in both dark matter~(DM) and baryonic physics impact galaxy formation. 
Integral to this effort is the development and training of Machine Learning models that have the capacity to identify subtle impacts from changing DM model assumptions in the face of uncertain baryonic physics.
The project will ultimately comprise thousands of simulations, spanning Cold Dark Matter~(CDM) and more complicated dark-sector models~\citep[for a review, see][]{2018Bertone} while simultaneously incorporating uncertainties in model predictions by varying baryonic feedback across the simulations. 
Different DM scenarios will be simulated for a range of targets---from clusters to ultra-faint satellites---providing an opportunity to study the impact of fundamental particle physics interactions on galactic and sub-galactic scales. 
This paper introduces \name and provides some case studies to demonstrate its use in machine learning applications.

The CDM paradigm has proven to be remarkably successful on large cosmological scales~\citep{2005SpringelA, 2010Reid, 2017Alam, 2018Springel}, and the current frontier is to test its validity on galactic and sub-galactic scales. 
In this paradigm, the DM is typically collisionless and interacts primarily through gravity.
However, recent theoretical work highlights that DM can live in a ``dark sector,'' with multiple particles and/or new forces that mediate non-gravitational interactions~\citep{2018Natur.562...51B}. 
Such dark-sector models can reproduce cosmological observables and yield the correct abundance of DM today, while simultaneously affecting the internal structure of galaxies as they evolve \citep[for a review, see][]{2017Arun}.

The potential degeneracy that arises from modifications to the internal structure of galaxies has led to an ongoing debate on several small-scale ``tensions'' where observations appear to conflict with CDM expectations \citep[for a review, see][]{2022Sales}. 
Classic examples include the diversity~\citep{2010Kuzio, 2015Oman}, missing-satellites~\citep{1999Moore, Klypin1999}, too-big-to-fail~\citep{2011Boylan, 2012Boylan, 2014Garrison, 2014Tollerud, 2015Papastergis}, satellite planes~\citep{2014Pawlowski, 2018Pawlowski}, and quiescent fractions~\citep{2019Hausammann, 2022Samuel} problems.
In some of these cases, the apparent tensions can be explained by an incomplete knowledge of baryonic feedback~\citep{2010Governato, 2013Brooks} and/or observational biases~\citep{2022Font, 2023Sawala}. 
For example, the missing-satellites problem, where many more satellites were found in cosmological simulations than observed around the Milky Way~(MW), is now considered solved after self-consistent stellar feedback was implemented in cosmological simulations~\citep{2019Garrison, 2021EnglerA} and observational completeness was accounted for \citep{2020Nadler}.

One of the key ingredients of \name is the treatment of astrophysical uncertainties that arise from baryonic feedback processes. 
Due to the limiting resolution of current hydrodynamic cosmological simulations, the astrophysics of stellar and black hole evolution is not emergent and must be implemented through phenomenological models.
While some models \citep[e.g.][]{FIRE2, 2019Marinacci} are progressing toward more explicit implementations of the physical processes that drive galaxy formation, any given ``sub-grid'' prescription depends on a set of parameters that remain uncertain, even if constrained to some extent by available observations. 
For a review of commonly used subgrid models and their components, see \cite{2020Vogelsberger}. 
Because energy injection due to Supernova~(SN) and Active Galactic Nuclei~(AGN) feedback plays an important role in redistributing the matter at the center of a halo \citep{2010Governato, 2011Teyssier, 2017Peirani, 2019Peirani, 2023Gebhardt}, large uncertainties in feedback modeling can translate into large uncertainties on certain DM properties. 
Accounting for these astrophysical systematics is therefore of critical importance to produce robust predictions for these properties. 
The DREAMS project will include thousands of simulations that span the uncertainty in alternative DM and galaxy formation models to catalyze the theory needed to utilize galaxies as particle physics laboratories.
With this dataset, \name will work toward the ultimate goal of testing CDM at the galaxy formation scale or discovering its alternate. 

Dedicated cosmological hydrodynamic simulations exist for a small number of CDM alternatives~\citep[e.g.][]{2014VogelsbergerB, 2017Robles, 2018LovellA, 2019Robertson, 2021Sameie, 2022Shen, 2022Kulkarni, 2023Rose, 2024Brown, 2024Correa}, but---due to their limited number---do not adequately account for cosmic variance, uncertainties in baryonic modeling, or the extent of available DM parameter space. 
The DREAMS project aims to adequately span this large multidimensional parameter space. 
Given the breadth of simulation data that will be available, the non-linear nature of structure formation, and the apparent degeneracies between the dark sector and baryonic physics, the project lends itself to statistical applications, such as machine learning. 
The CAMELS project~\citep{CAMELS}, which contains thousands of CDM simulations where astrophysical and cosmological parameters have been varied, provides a successful demonstration of the analyses that can ultimately be performed with DREAMS.
For example, \cite{2023Ni} showed that their neural networks can disentangle 28 simultaneously-varied parameters to make accurate constraints on cosmology and astrophysics.
Additionally, other works within the CAMELS project have shown that a variety of machine learning models can be used to constrain cosmological parameters while marginalizing over uncertainties in baryonic physics~\citep{2021Villaescusa, 2022Villanueva, 2022Villaescusa, 2023Santi}.
By exploring alternative DM models and extending the simulations down to smaller scales, \name will open new avenues of exploration for using machine learning methods to infer DM properties based on small-scale effects. 

\begin{table*}
    \centering
    \begin{tabular}{llllllll}
        \hline 
        Dark Matter        & \# of & Simulation   & Baryon       &  Cosmology & Mass Resolution & Spatial Resolution   & Notes  \\
        Prescription       & Sims  & Type         & Prescription &            & [$h^{-1} \Omega_\mathrm{m} \mathrm{M}_\odot$] & [$h^{-1}$~pc] & \\
        \hline
        Cold Dark Matter   & 1024  & Milky Way    & N-body       & Varied     & $4.0\times10^6$ & 305 & Forthcoming  \\
                                    & 1024  & Milky Way    & TNG       & Varied     & $4.0\times10^6$ & 305 &  Forthcoming \\
        \hline
        Warm Dark Matter   & 200   & Uniform Box  & N-body       & Fixed      & $3.3\times10^7$ & 1000 &  \cite{2024Rose} \\
                           & 1000  & Uniform Box  & N-body       & Varied     & $2.6\times10^8$ & 1000 &  \cite{2024Rose} \\
                           & 1024  & Uniform Box  & TNG          & Varied     & $2.6\times10^8$ & 1000  &  This paper\\
                           & 1024  & Milky Way    & TNG          & Fixed      & $4.0\times10^6$ & 305 &  This paper \\
        \hline     
    \end{tabular}
    \caption{List of simulation parameters for each suite in the DREAMS project. \textit{Col 1:} The name of the dark matter model included in the suite. \textit{Col 2:} The number of unique simulations in each suite, each with its own random initial density perturbations as well as dark matter and/or baryonic physics parameters~(if relevant). \textit{Col 3:} The type of simulation, either uniform box or Milky Way zoom-in.  \textit{Col 4:} The galaxy formation model that is included in the simulations, if any. \textit{Col 5:} Whether the cosmological parameters $\Omega_\mathrm{m}$ and $\sigma_8$ are varied within the suite. \textit{Col 6:} The dark matter mass resolution for each of the simulation suites divided by $\Omega_\mathrm{m}$. Note that dividing by $\Omega_\mathrm{m}$ makes the mass resolutions appear larger than those typically quoted in the literature. \textit{Col 7:} The gravitational softening for each simulation suite at $z=0$. The N-body WDM suites were studied in~\cite{2024Rose}, while the TNG WDM suites are the focus of this paper. The next suites planned as part of the DREAMS effort will focus on CDM, denoted above as ``forthcoming.''  Section~\ref{sec:discussion} discusses plans for other extensions to the DREAMS project.}
    \label{tab:simulations}
\end{table*}

This paper presents the first two hydrodynamical simulation suites that have been completed for the DREAMS project. For simplicity, we choose to start with thermal-relic Warm Dark Matter~\citep[WDM;][]{2000Hogan, 2001Sommer} where the initial matter power spectrum is suppressed below the free-streaming scale of the DM particle.
This model introduces only one additional simulation parameter (i.e., the WDM particle mass) relative to CDM.
The suppression in small-scale power makes WDM highly predictive. 
Current $2\sigma$ constraints from a joint analysis of strong gravitational lensing and satellite properties rule out masses below 9.7~keV~\citep{2021NadlerA}. 
Another joint analysis by \cite{2021Enzi}, which incorporates Ly$\alpha$-forest measurements, constrains WDM masses to $> 6.0$~keV.
Other constraints have also been set at 6.5~keV through observations of satellite properties~\citep{2021Nadler}, 5.6~keV through strong gravitational lensing~\citep{2020Hsueh}, and 5.3~keV through measurements of the Ly$\alpha$-forest~\citep{2017Irsic}.

We present results for both a uniform-box and MW zoom-in WDM simulation suite. 
Each suite contains 1024 distinct simulations over which parameters associated with SN and AGN feedback have been varied within the IllustrisTNG~\citep[TNG;][]{2018Pillepicha, 2018Weinberger} galaxy formation model. 
Notably, this is the largest zoom-in suite created to date for any DM model and provides a unique sample of MW-like classical satellites to study in a WDM cosmology.

We perform two separate case studies with these suites. 
The first utilizes emulation on the zoom-in suite to investigate how WDM and astrophysical parameters affect satellite counts around MW-mass hosts.
We find that the machine-learning-based analysis can disentangle how DM and astrophysics parameters independently affect satellite counts.
The second infers WDM particle properties from field-level data with CNNs using images of projected density fields from uniform-box simulations that vary DM, baryon, and cosmological properties. 
With this dataset, our analysis marginalizes over baryonic and cosmological uncertainties to infer the WDM particle mass from a range of astrophysical properties.

Moving forward, \name will grow to include additional suites for other DM models, including examples with multiple subspecies of DM and/or novel DM interactions. 
Within this framework, we aim to:
\begin{itemize}[
  align=left,
  leftmargin=\parindent,
  itemindent=0pt,
  labelsep=0.5pt,
  labelwidth=1em,
  itemsep=0.5em]
    \item understand the dependence of galaxy formation and galaxy properties on variations in the dark sector;
    \item identify new observables that can constrain properties of DM while accounting for uncertainties in cosmology and astrophysics;
    \item efficiently explore large swaths of dark-sector parameter space to set constraints and identify promising regions for detectability;
    \item train machine-learning models to infer DM properties while marginalizing over cosmological and astrophysical uncertainties.
\end{itemize}

This paper is organized as follows.
Section~\ref{sec:sim_methods} outlines the simulation design and specifications within the DREAMS project.
The next two sections, Sections~\ref{sec:emulation} and \ref{sec:inference}, outline the two different applications for the WDM suites that have been completed. 
Section~\ref{sec:discussion} outlines the possible extensions of this framework.
We conclude in Section~\ref{sec:conclusion}.
The appendices provide additional information for the methods and models included in this paper.
Appendix~\ref{app:zoom-in} outlines the procedure used to generate the zoom-in suite consisting of 1024 unique MW-mass galaxies.
Appendix~\ref{app:DM} details the different alternative DM models presented in this paper.
Finally, Appendix~\ref{app:ML} provides information on the machine learning architectures, training procedures, and data used in the analysis.

\section{Simulation Framework}
\label{sec:sim_methods}

The DREAMS project is comprised of separate suites of cosmological simulations--- either uniform boxes or zoom-ins.
Uniform boxes have the advantage of providing large statistics with millions of galaxies but at the cost of poor resolution.
Zoom-in simulations focus on fewer galaxies than uniform boxes but at much higher resolution for a similar computational cost.  
Including multiple simulation types allows one to focus on different observational probes of galactic structure, while the large number of simulations in each suite helps to capture cosmic variance.
Each hydrodynamical simulation in the suite also has a different set of baryonic feedback parameters.
These variations allow one to incorporate uncertainties in the astrophysical prescription into the final results, similar to what is done in CAMELS~\citep{CAMELS}.
Table~\ref{tab:simulations} summarizes the WDM hydrodynamical suites used in this paper, as well as currently available or forthcoming suites that are also part of the DREAMS project.

For the suites considered in this work, the simulations are advanced from $z=127$ to $z=0$ by the moving-mesh code \textsc{Arepo} \citep{Springel2010, Springel2019, Weinberger2020}.
\textsc{Arepo} uses a TreePM grid~\citep{Bagla2002} to solve gravity with a comoving spatial softening for all particles to reduce the numerical noise of close-body interactions.
Hydrodynamics are implemented using a pseudo-Lagrangian moving-mesh approach, which utilizes Voronoi tessellations around gas particles to fill the entire volume of the simulation.
These tessellations are adaptive to resolve a large range of spatial scales.
We generate 91 particle and group snapshots between $z=15$ and $z=0$.
The group snapshots include both \textsc{subfind}~\citep{2001Springel} and \textsc{rockstar}~\citep{2013Behroozia} galaxy catalogs with merger trees created using \textsc{sublink}~\citep{2015Rodriguez} and \textsc{consistent trees}~\citep{2013Behroozib}, respectively. 
For the analyses in this paper, we use the \textsc{subfind} catalogs, which define a galaxy using 3D overdensities and satellites as galaxies that are gravitationally bound to a larger halo.

The remainder of this section describes the implementation of the baryonic physics~(Sec.~\ref{sec:TNG}), cosmology~(Sec.~\ref{sec:cosmology}) and initial conditions~(Sec.~\ref{sec:initial}) in more detail. 

\subsection{Baryon Prescription}
\label{sec:TNG}

We employ the TNG galaxy formation model~\citep{2018Pillepicha, 2018Weinberger} for the two hydrodynamic simulation suites presented in this paper.
This version builds upon the earlier Illustris model~\citep{2013Vogelsberger, 2014Torrey} by implementing a kinetic AGN feedback and a self-consistent magnetohydrodynamic model.
Additionally, prescriptions for stellar formation and evolution, galactic winds, and galactic outflows have been updated from the Illustris model.
The TNG model has been tested across a wide range of environments and resolutions and reproduces several observational relations~\citep[e.g.,][]{2018Pillepichb,2018Springel, 2018Naiman, 2019Nelson,2019Marinacci}.
This subsection briefly reviews the relevant TNG prescriptions for this study---see \cite{2018Pillepicha} for a detailed discussion of the full TNG model.

Feedback prescriptions are one of the significant systematic uncertainties in hydrodynamical codes when modeling DM halo properties. 
In dwarf systems, for example, stellar feedback can redistribute the DM in the central regions of the galaxy, creating cored as opposed to cuspy density profiles depending on the mass of the host~\citep[e.g.,][]{2020MNRAS.497.2393L}.
The suites presented here adopt the same fiducial parameters that define the TNG model, except for three that specifically affect SN and AGN feedback. 
We vary parameters that adjust the strength of feedback over a wide range to both bracket uncertainties when analyzing a particular DM model and to avoid imposing restrictive priors on machine-learning models. 

The TNG subgrid model for SN feedback accounts for the ejection of winds, which move gas throughout a galaxy.  
There are two key inputs to this model: the mass-loading factor, $\eta_w$, and the stellar wind velocity, $v_w$. 
The former, which is directly related to the wind energy, is the ratio with which gas mass is lost through winds versus direct star formation. 
Updated from Illustris, the TNG model allows for a subdominant portion of the available energy from stellar feedback to be thermally released.
Specifically, the mass-loading factor is given by
\begin{equation}
    \label{eq:mass_loading}
    \eta_w = \frac{2}{v_w^2} e_w \left(1 - \tau_w \right) \, ,
\end{equation}
where $e_w$ is the specific energy available for generating winds in units of $10^{51}~\mathrm{erg}~\mathrm{M}_\odot^{-1}$, $v_w$ is the wind velocity (defined below), and $\tau_w$ is the fraction of energy that is released thermally.
While $\tau_w$ is fixed to its fiducial value of 0.1, we vary the specific energy logarithmically between $e_w \in [0.9, 14.4]$.
The fiducial value for $e_w$ in the TNG model is 3.6.

Increasing the strength of SN winds results in a decrease in the cosmic star formation rate density~(SFRD) for $z \gtrsim 1$ in the TNG model \citep{2018Pillepicha}. 
This feedback has the greatest effect on low-mass galaxies with $\mathrm{M}_* > 10^{10}~\mathrm{M_\odot}$, which comprise a significant portion of the SFRD for $z \gtrsim 2$~\citep{2014Torrey}.
At later times, AGN feedback can dominate changes to the cosmic SFRD by suppressing star formation in massive halos~\citep{2014Torrey}.
For a simulation of the same size and resolution as those presented in Section~\ref{sec:inference}, a $2\times$ increase in $e_w$ results in a lower SFRD at early times that is nearly consistent with observations~\citep{2018Pillepicha}.
The same increase to $e_w$ also reduces the stellar mass content of MW-mass ($\mathrm{M}_{\rm halo} \sim 10^{12}~\mathrm{M}_\odot$) below the observed values \citep{2018Pillepicha}.
Similar parameter variations within the Illustris model to those that we adopt here result in SFRDs that are nearly consistent with observations \citep{2013Vogelsberger, 2014VogelsbergerE}.
Within the TNG model, simultaneous variations of a $2\times$ increase in $e_w$ and an increased $v_w$ to reproduce the fiducial value of $\eta_w$ results in a SFRD that is too low across all redshifts \citep{2018Pillepicha}.
We choose to extend the parameter space of $e_w$ to $4\times$ the fiducial value so that extreme variations in this one parameter will cause the simulations to be inconsistent with observations and simultaneous variations with our other SN variable, discussed below, can result in the fiducial mass loading factor. 

The speed of the galactic winds created by SN feedback is scaled by the local DM velocity dispersion, $\sigma_{\scriptscriptstyle{\rm DM}}$, in a redshift-dependent manner:
\begin{align}
v_w = \mathrm{max} \left[\kappa_w \sigma_{\scriptscriptstyle{\rm DM}} \left(\frac{H_0}{H(z)} \right)^{1/3}, \, v_{w,\mathrm{min}} \right] \, ,
\end{align}
where $\kappa_w$ is a dimensionless normalization factor, $H(z)$ is the Hubble parameter, and $v_{w,\mathrm{min}}$ is the minimum wind speed. 
$\sigma_{\scriptscriptstyle{\rm DM}}$ is computed from the 64 nearest DM particles to the SN and $v_{w,\mathrm{min}} = 350~{\rm km/s}$.
We vary the dimensionless normalization factor logarithmically in the range $\kappa_w \in [3.7, 14.8]$.  The fiducial value for $\kappa_w$ in the TNG model is 7.4. 

Increasing the speed of the SN winds, $\kappa_w$, results in a decrease in the cosmic SFRD for $z \lesssim 4$ where the most extreme variation is inconsistent with observations below $z \sim 3$~\citep{2018Pillepicha}. 
Similar to a $2\times$ increase in $e_w$, increasing $\kappa_w$ by $2\times$ results in a decrease in the stellar mass content of MW-mass galaxies that are inconsistent with observations \citep{2018Pillepicha}.
Similar parameter variations in the Illustris model to those that we adopt here result in too low SFRD at late times for $2\times$ their fiducial value but are consistent with observations for $0.5\times$ their fiducial value~\citep{2013Vogelsberger, 2014VogelsbergerE}. 
In the same spirit as the $e_w$ parameter, we adopt a broad range for $\kappa_w$.

The overall strength of AGN feedback can also play an important role in shaping the DM halos around galaxies \citep{2022Anbajagane}. 
In the TNG model, a black hole with a mass of $8 \times 10^{5}~h^{-1}~\mathrm{M}_\odot$ is seeded at the center of a halo that reaches a mass of $5 \times 10^{10}~h^\mathrm{-1} \mathrm{M}_\odot$.  
Accretion onto the black hole is modeled by a spherical Bondi-Hoyle model~\citep{1944Bondi} and feedback is implemented for two phenomenologically different states. 
The low-accretion state corresponds to a system where hot gas accretes isotropically onto the black hole~\citep{1976ApJ...204..187S, 1977ApJ...214..840I}, while the high-accretion state corresponds to cooler gas that accretes through a disk~\citep{1973A&A....24..337S}.  
The high-accretion state dominates for galaxies with stellar mass $\lesssim 10^{10.5}~\mathrm{M}_{\odot}$, while the low-accretion state takes over at higher stellar masses~\citep{2017Weinberger}.
Therefore, for the MW zoom-in suite, the high-accretion state is the more relevant of the two. 
This is also true for the uniform-box suite, where the number of halos with stellar mass greater than $\sim 10^{10.5}~\mathrm{M}_{\odot}$ is typically $\sim 40$ and the number of halos less than $\sim 10^{10.5}~\mathrm{M}_{\odot}$ with a black hole is typically $\sim 10^3$ per simulation.
For this reason, we only vary the high-accretion state here.

In this state, the feedback energy released from accretion is given by 
\begin{equation}
    \Delta \dot{E} = \epsilon_{f,\mathrm{high}} \epsilon_r \dot{M}_{\scriptscriptstyle{\rm BH}} c^2 \, ,
\end{equation}
where $\epsilon_{f,\mathrm{high}}$ is the fraction of energy that is transferred to the nearby gas,
$\epsilon_r$ is the radiative efficiency, and $\dot{M}_{\scriptscriptstyle{\rm BH}}$ is the black hole accretion rate. 
All parameters are kept fixed to their TNG fiducial values, except for $\epsilon_{f,\mathrm{high}}$, which is varied logarithmically in the range $\epsilon_{f,\mathrm{high}} \in [0.025, 0.4]$. 
Its TNG fiducial value is 0.1.\footnote{The AGN parameter we vary differs from the two parameters used in the CAMELS project, which were found to induce negligible differences across many observables~\citep{CAMELS}.}
We chose to vary $\epsilon_{f,\mathrm{high}}$ as it is a tuned parameter in the TNG model.

Increasing the strength of the AGN feedback, $\epsilon_{f,\mathrm{high}}$, results in an increase in the cosmic SFRD and a corresponding increase in the stellar-mass-halo-mass relation for $z \gtrsim 3$~\citep{2023Ni}.
A $4\times$ increase in $\epsilon_{f,\mathrm{high}}$  results in a $\sim 3 \times$ increase in the SFRD at late times for similar simulations to those presented in Section~\ref{sec:inference}~\citep{2023Ni}.
This variation is slightly larger than observational uncertainties on the cosmic SFRD~\citep{2013Behroozi}.
In \cite{2017Weinberger}, $\epsilon_{f,\mathrm{high}}$ is varied by a factor of four and shown to affect the gas fraction and star formation rate, which still remain within the bounds of observed results.
The radiative efficiency, proportional to $\epsilon_{f,\mathrm{high}}$ in our simulations, has also been shown to depend on mass, suggesting that variations around the fiducial parameter would be needed~\citep{Davis_2011}.

A number of studies have investigated how varying these astrophysical parameters can affect various galactic properties~\citep{2014VogelsbergerE, 2014Torrey, 2017Weinberger, 2018Pillepicha, 2021Roca, 2023Ni}.
We adopt parameter ranges that when varied individually are consistent with observations for the majority of the range that we include.
The high end of our two SN parameters, $e_w$ and $\kappa_w$, are shown to be inconsistent with observations when varied individually \citep{2018Pillepicha}.
Only a single instance of simultaneous parameter variations is available, an increase in both $e_w$ and $\kappa_w$, and is shown to be inconsistent with observations across all redshifts \citep{2018Pillepicha}.
Overall, the extrema of the parameter ranges we adopt are likely to fall outside current observational uncertainties, but are still useful to mitigate prior effects.

Except for the two SN parameters and one AGN parameter described above, all other parameters in the TNG model are fixed across the simulation suites.
The uniform boxes we present have a similar resolution to the TNG300-1 box, and the zoom-in simulations are at $\sim 2 \times$ lower mass resolution than TNG50-1.

\subsection{Cosmology}
\label{sec:cosmology}

For a given simulation suite, the cosmological parameters can either be fixed or varied.  
In cases where they are fixed, we assume a cosmology consistent with \cite{2016Planck}, where $\Omega_\mathrm{m}$ = 0.302, $\Omega_\Lambda = 0.698$, $\sigma_8 = 0.839$ and $H_0 = 100~h~{\rm km/s}$ such that $h = 0.691$.
We fix $\Omega_{\rm b}$ at 0.046 for the hydrodynamic simulations and 0.0 for the N-body simulations.

In other instances, we vary $\Omega_{\rm m}$ and $\sigma_8$ within the following ranges: 
$0.1 < \Omega_{\rm m} < 0.5$ and $0.6 < \sigma_8 < 1.0$.
All other cosmological parameters are kept fixed.  
These ranges purposefully exceed the current uncertainties from CMB analyses~\citep{2020Planck} so that
the prior distribution has minimal impact on machine learning models.
These parameter ranges are consistent with those used by the CAMELS project \citep{CAMELS}, which has shown that machine learning models can infer cosmological values accurately across the entire range of each parameter~\citep{CMD}.
Cosmological and astrophysical variations, when present, are varied according to a Sobol sequence~\citep{sobol}.

\subsection{Initial Conditions}
\label{sec:initial}
 \subsubsection{Uniform-Box Simulations}
\label{sec:uniform}

Uniform-box simulations cover a large periodic volume at one resolution and have the advantage of including thousands to millions of galaxies across a wide range of mass and size scales. 
These simulations provide a cosmological context that ensures a realistic environment for galaxy formation and evolution.
Moreover, the large sample size enables the study of galaxy properties that require substantial statistics to disentangle.
For example, utilizing information from thousands of galaxies, one can explore whether a large sample of galactic systems can be synthesized to uncover DM models that only induce subtle changes to individual galaxy properties.  

Despite these advantages, state-of-the-art uniform-box simulations are typically not able to resolve structures with a stellar mass below $\sim 10^7~\mathrm{M}_\odot$~\citep{firebox, 2019Nelson, 2019Pillepich}.
While this is not necessarily a concern for investigations into large-scale structure and cosmology, it can be problematic if one wants to study DM models where deviations from CDM occur on sub-galactic scales.

\begin{figure*}
    \centering
    \includegraphics[width=\textwidth]{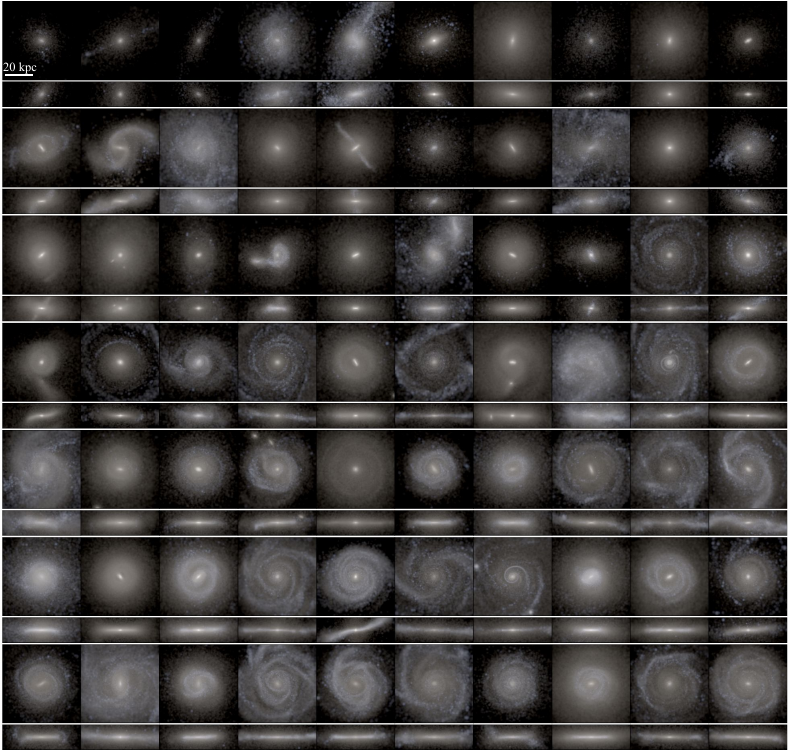}
    \caption{Three-dimensional stellar light reconstructions of the central galaxies from a small sample~($\sim7\%$) of the WDM zoom-in suite that includes variations to the DM particle mass and the TNG galaxy formation model. 
    The images are ordered (from top left to bottom right) by increasing disk-to-stellar-mass ratios and consist of a wide range of systems ranging from red spherical to blue disk-dominated galaxies.
    The images are made from the \textit{i, r, g} Sloan broadband filters. Each image shows a $50\times 50 \times 15~h^{-3}~{\rm kpc^3}$ volume. Each volume is projected along the short axis for the face-on images or along one of the long axes for the edge-on images.}
    \label{fig:zoom_images}
\end{figure*}

The DREAMS project does include several suites that are cosmological boxes, denoted `Uniform Box' in Tab.~\ref{tab:simulations}.
Each of these simulations covers a $(25~h^{-1}~{\rm Mpc})^3$ volume with periodic boundary conditions and starts from a $2 \times 256^3$ grid of DM and gas particles.
The DM mass resolution for these simulations is $7.81 \times (\Omega_{\rm m}/0.302) \times 10^{7}~h^{-1} \mathrm{M}_\odot$. 
The baryon mass resolution, corresponding to the average mass of the star particles throughout the simulation, is $1.27\times 10^7~h^{-1} \mathrm{M}_\odot$.
The spatial resolution is determined by the gravitational softening factor that is applied to reduce numerical noise.
The uniform-box simulations have a spatial resolution of $1.0~h^{-1}~{\rm kpc}$ at redshift $z=0$.

The initial conditions for each simulation are created with \textsc{Ngenic}~\citep{Springel2019}.
\textsc{Ngenic} begins with a glass distribution of particles that evolves according to the equations for second-order Lagrangian perturbation theory up to $z=127$.
The glass distribution is convolved with Gaussian random noise to create a unique distribution of matter from which to start the simulations.
Each simulation in \name is given a different random seed to create the Gaussian random noise field ensuring that each simulation starts from a unique distribution of DM perturbations.

\subsubsection{Zoom-in Simulations}
\label{sec:zooms}

We also introduce simulations that target MW-mass galaxies at much higher resolution.
The advantage of these so-called ``zoom-in'' simulations is that they can better resolve the inner structure and dynamics of a small sample of galaxies for a reduced computational cost. 
This more detailed view of the internal dynamics of a MW system is imperative for testing deviations from CDM.

For our first suite of zoom-in simulations, we focus on MW-mass galaxies.
Currently, the largest sample of high-resolution MW-mass galaxies is the TNG50 simulation~\citep{2019Nelson, 2019Pillepich}.
The TNG50 simulation contains 198 MW-mass galaxies with a DM mass resolution of $6.5 \times 10^5 ~ h^{-1}~\mathrm{M}_\odot$ and a baryon resolution of $1.2 \times 10^5 ~h^{-1}~\mathrm{M}_\odot$.
The zoom-in suites in \name include $\sim$5$\times$ as many MW-mass galaxies at $2\times$ lower resolution while also varying the DM and astrophysical prescriptions between the simulations.  
These are not direct MW-analogs as they are chosen to not be part of a local group that includes a nearby massive neighbor (for computational reasons).  
Future work will explore possibilities of extending the suite to include additional environments with analogs of Andromeda, M31.

We use \textsc{music}~\citep{2011Hahn} to generate the initial conditions for the zoom-in simulations in the suite.
\textsc{music} works similarly to \textsc{ngenic} in that it first convolves a uniform particle distribution with a Gaussian random field and then uses second-order Lagrangian perturbation equations to evolve the particles to $z=127$.
Each MW-mass target is chosen at random from a different $100~h^{-1}~\mathrm{Mpc}$ box with an isolation criterion of no neighbors more massive than $7.2\times10^{11}~h^{-1}~\mathrm{M}_\odot$ within $1~h^{-1}~\mathrm{Mpc}$.
The high-resolution regions of the zoom-in simulations presented here have a DM mass resolution of $1.2 \times 10^6~h^{-1}~\mathrm{M}_\odot$ 
and a spatial resolution of $0.31~h^{-1}~{\rm kpc}$ at $z=0$.
The average baryon mass resolution is $1.9 \times 10^5~h^{-1}~\mathrm{M}_\odot$.
Further details on the multi-step procedure required to create the initial conditions for the zoom-in simulations are provided in Appendix~\ref{app:zoom-in}. 

\section{Emulating WDM Satellite Counts}
\label{sec:emulation}

This section introduces the zoom-in suite focused on WDM MW-mass galaxies. 
This is the largest sample (to date) of MW-mass galaxies that have been simulated with a realistic galaxy formation model at high resolution.
WDM differs from CDM by including free-streaming velocities that suppress early-universe density perturbations at small scales because the DM escapes from potential wells~\citep{2001Bode}.
We consider thermal-relic WDM where the initial velocities correspond to the time at which the particles decouple from the surrounding plasma.
Although non-thermal production mechanisms for WDM are also possible~\citep{Banerjee:2023utz}, the thermal scenario provides a convenient starting point for its simplicity: the only relevant DM model parameter is the fundamental WDM particle mass.  

\subsection{WDM Zoom-In Suite}

The WDM zoom-in suite consists of 1024 unique MW-mass galaxies at a baryon~(DM) mass resolution of $1.9 \times 10^5~(1.2\times 10^6)~h^{-1}~\mathrm{M}_\odot$.\footnote{This value is different than the one quoted in Tab.~\ref{tab:simulations} because $\Omega_\mathrm{m}$ is now fixed at 0.301712.} In generating this suite, the baryonic parameters~(two SN and one AGN) are varied, but the cosmological parameters are kept fixed.  Additionally, 
the WDM particle mass is varied uniformly in $\mathrm{M_{\scriptscriptstyle{WDM}}^{-1}}$ between 0.033 and 0.555 ${\rm keV^{-1}}$ (1.8--30~keV).  
For an overview of the WDM implementation in DREAMS and a discussion of the impact of artificial clumping, see Appendix~\ref{app:DM}.   

\begin{figure}
    \centering
    \includegraphics[width=\columnwidth]{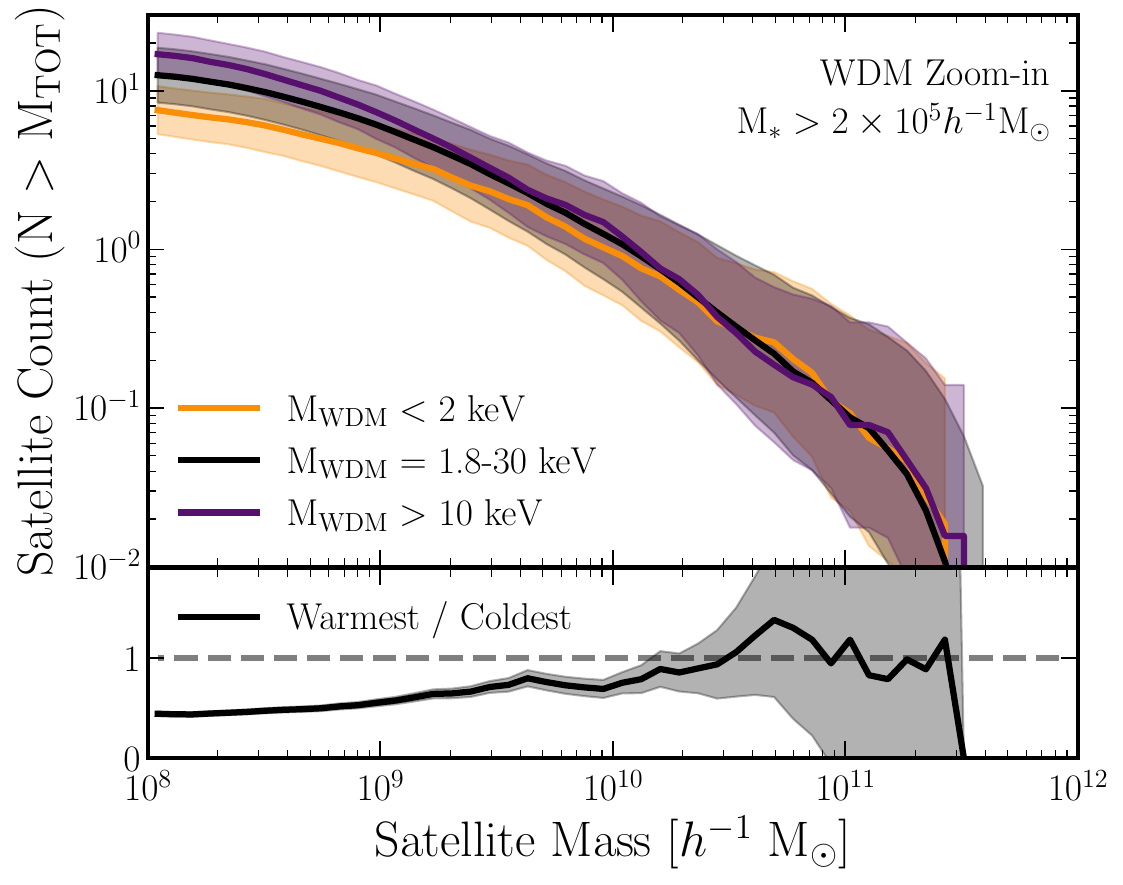}
    \caption{The satellite mass function for galaxies in the zoom-in WDM suite, which is discussed in Section~\ref{sec:emulation}.  We include all satellites with $\mathrm{M_{halo}}~>~10^8~h^{-1}~\mathrm{M_\odot}$ and $\mathrm{M_*}~>~2~\times~10^5~h^{-1}~\mathrm{M_\odot}$, which are within 300~kpc of the central Milky Way-mass galaxy. The data is subdivided into two additional subgroups according to WDM particle mass: M$_{\scriptscriptstyle{\rm WDM}}<2~{\rm keV}$~(orange) and M$_{\scriptscriptstyle{\rm WDM}}>10~{\rm keV}$~(purple).  Each subset contains $\sim 100$ simulations.  For each case, the average counts are shown by the solid line and the average 1$\sigma$ deviation is shown as the shaded band. The bottom panel shows the ratio between the low-WDM-mass~(orange) and high-WDM-mass~(purple) subsets. The spread in the ratio is indicated by the shaded band and becomes large above $\sim 2 \times 10^{10}~\mathrm{M}_\odot$ due to the small number of satellites in this mass range. 
    }
    \label{fig:satellite_stats}
\end{figure}

For the galaxies in the suite, the central halo mass has a mean and standard deviation of $(8.8 \pm 1.2) \times 10^{11}~h^{-1}~\mathrm{M}_\odot$.  The host disk properties exhibit considerable variation in morphology due to the spread in the baryonic physics parameters.  For example, the stellar mass of the disk\footnote{To calculate the disk mass, we first calculate the angular momentum about the system's eigenvector of the moment of inertia tensor that aligns with the principle angular momentum axis for each star particle within 50~kpc of the host.
Any material rotating with at least 70\% of the maximum angular momentum is designated as being part of the disk.} has a mean and standard deviation of $(4.1\pm 2.8) \times10^{10}~h^{-1}~\mathrm{M}_\odot$. Additionally, one can consider the disk-to-total-mass ratio~($\rm{D/T}$), which quantifies the ratio of co-rotating to total stellar mass within 50~kpc of the host's center. Across all simulations in the suite, the central galaxies have an average {\rm D/T} and 16-84\% spread of $0.43^{+.20}_{-.22}$.
This average aligns well with other simulated MW-mass halos, but with a larger range likely due to the parameter variations and large sample size used here~\citep{2016Obreja, 2017Grand}.
Roughly 40\% of the galaxies have a disk-to-total-mass ratio greater than 0.5. 
Current estimates of the MW's D/T ratio are near 0.8, toward the higher end of ratios exhibited in this suite~\citep{2020Cautun}.

\begin{figure*}
    \centering
    \includegraphics[width=.98\columnwidth]{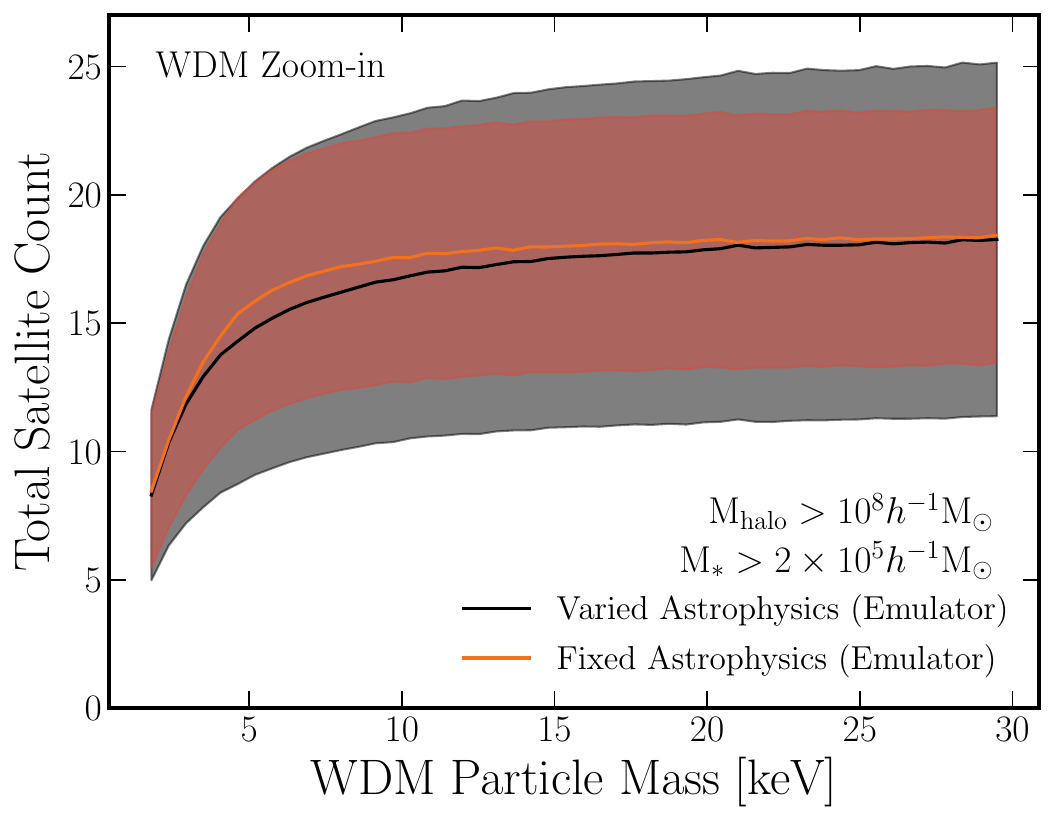}
    \includegraphics[width=.98\columnwidth]{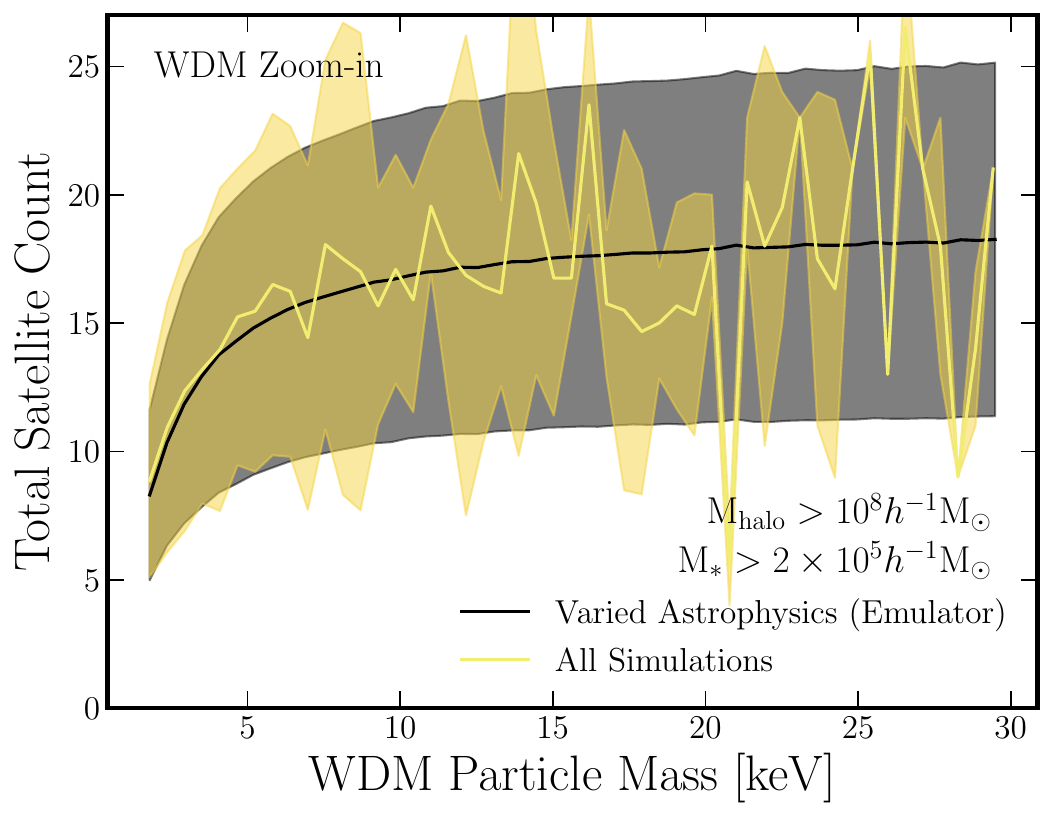}
    \caption{Emulator results from the neural network trained on the MW zoom-in simulation suite with varied astrophysics and WDM particle mass.
\emph{Left:} The number of satellites with $\mathrm{M_{halo}}~>~10^8~h^{-1}~\mathrm{M_\odot}$ and $\mathrm{M_{*}}~>~2~\times~10^5~h^{-1}~\mathrm{M_\odot}$ around a MW-mass galaxy as a function of WDM particle mass for the fiducial astrophysical parameters~(orange) and varied astrophysical parameters~(black). 
    The solid line represents the mean value across all galaxies in the sample with the shaded region covering the $\pm 1\sigma$ spread. 
    The uncertainty on the fixed-astrophysics results~(orange) represents the expected spread from cosmic variance.
    The uncertainty on the varied-astrophysics results~(black) represents the combined uncertainty from cosmic variance and uncertainty from SN and AGN feedback modeling.
    \emph{Right:} The total number of satellites around a MW-mass galaxy predicted from the simulations~(orange) and emulator~(black) with the same satellite selection as the left panel.
    Simulations are grouped in 0.5~keV bins including all astrophysics variations.
    Emulator results for both panels are sampled at 0.05~keV intervals and then smoothed with a top-hat kernel with a width of 0.5~keV to match the bin size of the simulation results.
    }
    \label{fig:emulator_WDM}
\end{figure*}

Figure~\ref{fig:zoom_images} presents a small sample~($\sim7\%$) of MW-mass galaxies from the WDM zoom-in suite depicting face-on and edge-on views.
The galaxies are rotated so that the angular momentum, computed using all star particles within 10~kpc of the potential minimum, points toward the viewer.
The images are then ordered (from top left to bottom right) by increasing D/T ratio.  The three-color image is comprised of the Sloan $g$, $r$, and $i$ filters as the blue, green, and red components of the image, respectively, with no correction for dust extinction.  Overall, the population of MW galaxies in this suite covers a wide range of morphologies from red spherical galaxies to blue disk-dominated systems. 

\subsection{Emulator Results}
\label{sec:emulator_results}

As a first application of the WDM zoom-in suite, this section explores how the number of satellites around a MW-mass halo depends on the WDM particle mass, as well as cosmic variance and other astrophysical uncertainties. 
This investigation includes all satellites that have a total DM halo mass of at least $10^8~h^{-1}~\mathrm{M}_\odot$ and a stellar mass of at least $2\times10^5~h^{-1}~\mathrm{M}_\odot$. All satellites must also be within 300~kpc of the central galaxy.
Since star formation is inefficient in low-mass halos (compared to $10^{10}$--$10^{13}~h^{-1}~\mathrm{M}_\odot$ halos), most galaxies in the suite host $\mathcal{O}(10)$ dark satellites that are not included in this analysis. 

The top panel of Fig.~\ref{fig:satellite_stats} shows the satellite mass function averaged over all MW-mass galaxies in the zoom-in suite~(black).
We also split the dataset into two subsets that include the $\sim100$ warmest (M$_{\scriptscriptstyle{\rm WDM}}<2~{\rm keV}$; orange) and coldest (M$_{\scriptscriptstyle{\rm WDM}}>10~{\rm keV}$; purple) WDM models in the suite.
The solid line corresponds to the average satellite mass function from each subset and the bands correspond to the $\pm 1\sigma$ spread, encompassing variations in host merger history, the WDM particle mass, and the three free astrophysical parameters.
The bottom panel of Fig.~\ref{fig:satellite_stats} shows the ratio of warmest~(orange) to coldest~(purple) subsets where the solid line shows the ratio of the averages from the top panel and the shaded region represents the $\pm 1\sigma$ spread about the average.

Lower WDM particle masses lead to a clear effect on the number of satellites around a MW-mass galaxy.
For $\sim 10^{10}~\mathrm{M}_\odot$ galaxies, there is a $\sim 30\%$ suppression from the coldest to warmest subgroups.
For less massive satellites, down to $\sim 10^{8}~\mathrm{M}_\odot$, there is a $\sim50\%$ reduction in the number of satellites.
Even accounting for variance from the spread in astrophysical parameters, there is a clear separation between the warmest and coldest datasets for galaxies less massive than $\sim 10^{10}~h^{-1}~\mathrm{M}_\odot$, as is expected for these WDM masses~\citep{2014Lovell, 2017Read}.

In total, the suite of WDM zoom-ins contains  $\sim 15,200$ luminous satellite galaxies and this large dataset can be used as a training set for a variety of machine learning applications. 
For this example, we build a neural network emulator that acts as a multi-dimensional interpolator, taking as input the simulation parameters and outputting the number of satellites around a MW-mass halo.
This method can be used to explore how a variety of astrophysical features, such as star formation rates, bar sizes, satellite mass functions, etc. change with each of the four parameters varied in this simulation suite.
While limited to interpolating within the parameter range provided, these emulators provide a useful tool for understanding complex and multidimensional trends within the data.
See Appendix \ref{app:emulator} for details on the architecture, training, and validation.

Figure~\ref{fig:emulator_WDM} shows the number of luminous satellites with $\mathrm{M_{halo}} > 10^8~h^{-1}~\mathrm{M}_\odot$ and $\mathrm{M_*}~>~2~\times~10^5~h^{-1}~\mathrm{M}_\odot$ around the MW-mass host as a function of the WDM particle mass in keV. 
In the left panel, emulator results are shown for the number of satellites when only the WDM particle mass is varied (orange) and when all three astrophysical parameters ($\epsilon_w$, $\kappa_w$, and $\epsilon_{f, {\rm high}}$) are also varied (black).
For the fiducial astrophysics results, the emulator is sampled every 0.05~keV across the range of WDM particle masses. 
The results are then smoothed with a top-hat filter with a kernel width of 0.5~keV.
For the varied astrophysical results, the emulator is sampled $10^3$ times every 0.05~keV where each of the 100 samples has a unique set of astrophysical parameters.
The results for the varied astrophysics model are then smoothed with a top-hat filter with a kernel width of 0.5~keV.

In Fig.~\ref{fig:emulator_WDM}, each solid line corresponds to the average number of satellites at a given WDM mass, and the shaded region covers the $\pm 1 \sigma$ spread in the data.
For the ``Fixed Astrophysics'' results, the spread corresponds to cosmic variance. 
At 30~keV, the coldest model, the emulator predicts an average of $16.9~\pm~4.9$ satellites around the host.
Decreasing the WDM particle mass has little effect on the number of satellites down to M$_{\scriptscriptstyle{\rm WDM}} \sim 10$~keV. 
Below this point, the number of satellites diminishes quickly.
At M$_{\scriptscriptstyle{\rm WDM}} \sim 1.8$~keV, the emulator predicts an average of $9.3~\pm~2.9$ satellites around an isolated MW-mass galaxy.

For the ``Varied Astrophysics'' results, the shaded band corresponds to both the cosmic variance as well as the uncertainty in the astrophysics parameters.
A M$_{\scriptscriptstyle{\rm WDM}} = 1.8~(30)$~keV model results in a satellite count of $8.7~\pm~3.3$~($16.7~\pm~7.1)$.
Similar to the fiducial astrophysics results, the emulator predicts little suppression from decreasing WDM particle mass down to 10~keV.

Across the WDM particle masses investigated here, including astrophysical variations results in minimal changes to the average number of satellites predicted around an isolated MW-mass galaxy.
However, including astrophysics uncertainties increases the spread in the number of satellites from an ensemble of MW-mass galaxies.
The upper bound on the uncertainty band is minimally affected up to WDM masses of 6~keV, after which variations on the SN and AGN feedback parameters increase the spread.
The lower bound on the uncertainty band increases faster, after WDM masses of $\sim 2.5$~keV.

In the right panel, the emulator results for the ``Varied Astrophysics'' dataset are shown again~(black), along with the binned simulation results~(orange).
The simulation results are binned in 0.5~keV bins with the same selection criteria as the emulator results.
The average and spread in the number of satellites agree well between the simulation and emulator results at low WDM mass.
At high WDM mass, the number of simulations per bin decreases due to the uniform sampling in $\mathrm{M_{\scriptscriptstyle{\rm WDM}}^{-1}}$.
Within this range, the simulation results become more stochastic, but the emulator results show a smooth relation across all WDM particle masses.

\begin{figure*}
    \centering
    \includegraphics[width=\textwidth]{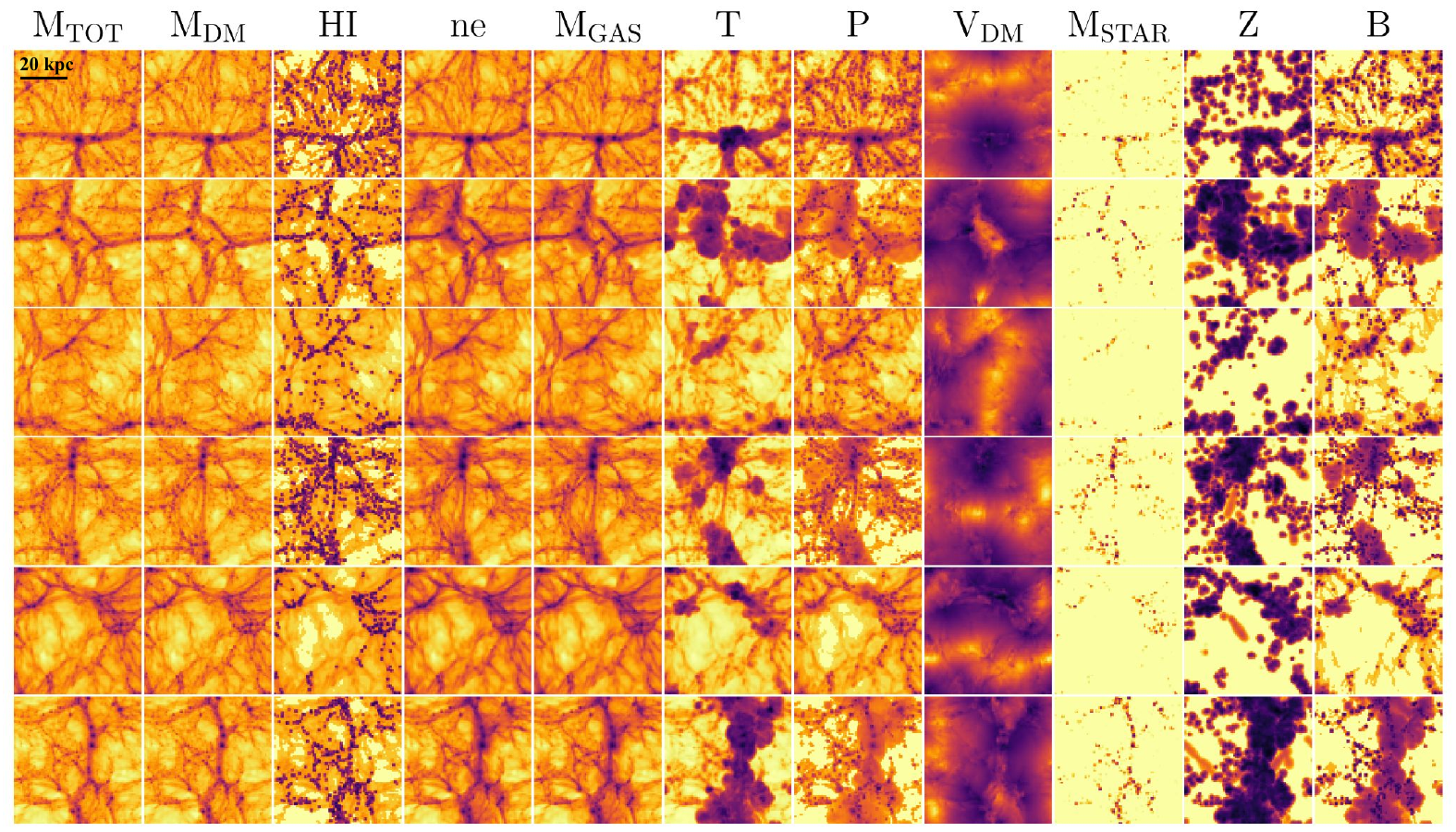}
    \caption{Image slices from the WDM uniform-box suite with varied cosmology and TNG physics. Each column represents a three-dimensional reconstruction of the given property that is then projected onto two dimensions: 
    total matter density, M$_\mathrm{TOT}$;
    dark matter density, M$_\mathrm{DM}$; 
    neutral hydrogen gas density, HI; 
    electron number density, ne;
    baryonic gas density, M$_\mathrm{GAS}$;
    gas temperature, T; 
    gas pressure, P;
    dark matter velocity modulus (speed), V$_\mathrm{DM}$;
    stellar mass density, M$_\mathrm{STAR}$;
    gas metallicity, Z;
    and magnetic field strength, B.
    Each row represents a projection from a different simulation taken randomly from the suite. Each image covers a $25\times 25 \times 5~h^{-1}~{\rm Mpc}$ volume projected along the short axis.}
    \label{fig:box_images}
\end{figure*}

Investigating how each astrophysical parameter affects the number of satellites around a MW-mass host reveals that $\epsilon_w$, which is related to the SN wind energy, has the largest impact on the number of satellites across all WDM particle masses. 
Across the range of $\epsilon_w \in [0.9, 14.4]$, the average number of satellites for a 30 keV WDM model ranges from 28 (low $\epsilon_w$) to 12 (high $\epsilon_w$) satellites.
The parameter with the second-most significant impact is $\kappa_w$, which is related to the SN wind velocity.
Across the range of $\kappa_w \ in [3.7, 14.8]$, the average number of satellites for a 30 keV WDM model ranges from 15 (low $\kappa_w$) to 20 (high $\kappa_w$).
The parameter that varies the AGN feedback, $\epsilon_{f, {\rm high}}$, has essentially no effect across the range that we vary.

Equation~\ref{eq:mass_loading} defines the mass loading factor, which relates $\epsilon_w$ and $\kappa_w$ and determines how much mass each SN will expel from a satellite galaxy.
Based on this equation, one might expect the wind velocities to have a greater impact on the number of satellites than the wind energy, since $\eta$ scales as $\epsilon_w$ versus $\kappa_w^{-2}$. 
This intuition is inconsistent with the emulator results, however.  
One possible explanation is that the wind velocities have a minimum allowed value $v_{w,\mathrm{min}}$.
For low-mass systems, the wind velocity may be capped at this value, even given the variations on $\kappa_w$.  
Additionally, at higher wind speeds, the gas ejected from the SN has less time to interact with surrounding material, which can further dampen the effect.

Unsurprisingly, the variations in AGN feedback do not affect the number of satellites around a MW-mass galaxy because nearly all the satellites are not massive enough to support a black hole that produces significant feedback. 
Similarly, the super-massive black hole in the central galaxy does not produce enough feedback to regularly disrupt surrounding satellites.
While it is possible that a satellite crossing through the jet of an AGN could be disrupted, this does not appear to be a significant factor around the MW-mass systems studied here.


\begin{figure}
    \centering
    \includegraphics[width=\columnwidth]{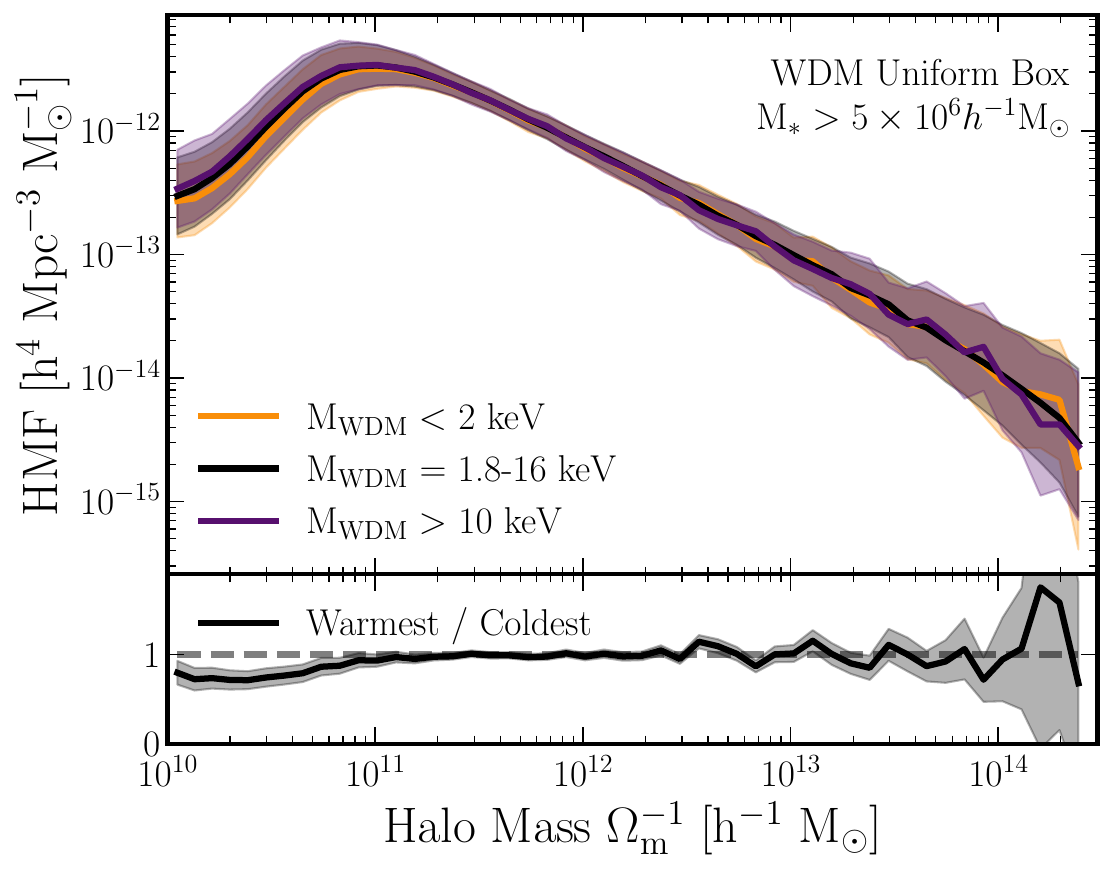}
    \caption{The halo mass function~(HMF) for the uniform-box WDM suite presented in Section~\ref{sec:inference}. We include all halos with $\mathrm{M_{halo}}~>~10^{10}~h^{-1}~\mathrm{M_\odot}$ and $\mathrm{M_*}~>~5~\times~10^6~h^{-1}~\mathrm{M_\odot}$. The data is also subdivided into two subgroups according to WDM particle mass: M$_{\scriptscriptstyle{\rm WDM}}< 2$~keV~(orange) and M$_{\scriptscriptstyle{\rm WDM}}>10$~keV~(purple). Each subset contains $\sim 100$ simulations.  For each case, the average counts are shown as solid lines and the average $\pm1\sigma$ deviation is shown as the shaded band. The bottom panel shows the ratio between the low-WDM-mass~(orange) and high-WDM-mass~(purple) subgroups. The spread in the ratio is large above $\sim 10^{13}~\mathrm{M}_\odot$ due to the small sample of halos in this mass range. 
    }
    \label{fig:box_stats}
\end{figure}

\section{Predicting WDM Mass from Galaxies}
\label{sec:inference}

This section explores the WDM uniform-box TNG suite in more detail.
It consists of 1024 hydrodynamic cosmological simulations, each with different cosmology, astrophysics, initial density field, and WDM particle mass.
Each row of Fig.~\ref{fig:box_images} shows a $25 \times 25 \times 5~h^{-3}~{\rm Mpc}^3$ slice from five random simulations at $z=0$.   
The columns correspond to a sample of properties that are tracked for each simulation, which are described in the figure caption.
See Appendix~\ref{app:CNN} for a description of how these images are created and a more detailed description of the different fields displayed in Fig.~\ref{fig:box_images}.

So far, constraints on the WDM particle mass at $z=0$ have largely come from observations of satellite properties~\citep{2021Nadler}, with other constraints coming from higher redshifts through the Ly$\alpha$ forest~\citep{2017Irsic}, strong lensing~\citep{Gilman:2019nap,2020Hsueh}, and UV luminosity functions~\citep{Corasaniti:2016epp,Sabti:2021unj}.
Each of these constraints relies on various summary statistics believed to capture information that constrains the WDM particle mass.
Recent investigations have produced better constraints by combining multiple statistics into one joint constraint~\citep{2021NadlerA, 2021Enzi}.

This section considers the constraining power of field-level data that is based on the DREAMS uniform-box suite.  Specifically, the data consists of 2D density projections of a $25~\times~25~\times~5~h^{-3}~{\rm Mpc}^3$ volume and a mass resolution of $7.81~\times~(\Omega_{\rm m}/0.302) \times 10^{7}~h^{-1} \mathrm{M}_\odot$.  The analysis does not rely on any chosen summary statistic, and the results are only limited by the amount and quality of the data given. While the example in this section is illustrative of the general technique, more sophisticated iterations in the future can generate mock data catalogs based on the DREAMS suites to make concrete predictions for any observable of interest. This will enable one to understand the constraining power of a given observable in testing a particular DM model while marginalizing over uncertainties from astrophysics and cosmology.

\subsection{General WDM Features Across Simulations}
\label{sec:wdm_stats}

To build intuition for the content of the WDM uniform-box suite, we begin with a general discussion of the simulated galaxies. 
Figure~\ref{fig:box_stats} shows the halo mass function~(HMF) for all halos more massive than $10^{10}~h^{-1}~\mathrm{M}_\odot$, the lowest-mass resolved in these simulations, with a stellar mass of at least $5 \times 10^{6}~h^{-1}~\mathrm{M}_\odot$. The turn-over at $10^{11}~h^{-1}~\mathrm{M}_\odot$ arises because lower-mass halos have a greater chance of not meeting the stellar mass cut that is applied.  As in Fig.~\ref{fig:satellite_stats}, the results are provided for all WDM masses~(black) as well as the coldest subset with M$_{\scriptscriptstyle{\rm WDM}} > 10$~keV~(purple) and the warmest subset with M$_{\scriptscriptstyle{\rm WDM}} < 2$~keV~(orange).  Solid lines denote the average halo counts, whereas the bands denote the $\pm 1\sigma$ spread arising from cosmic variance as well as uncertainties in astrophysics and cosmological parameters.   

As expected, the lowest-mass halos near $10^{10}~h^{-1}~\mathrm{M}_\odot$ have the strongest suppression ($\sim 30\%$), on average, between the warmest and coldest models.
Any signal for the highest-mass galaxies is dominated by cosmic variance as these simulations lack a large number of groups and clusters.
However, WDM is not expected to impact halo abundance at these scales.\footnote{We have also compared the stellar-mass-to-halo-mass~(SMHM) relation for all host halos more massive than $10^{10}~h^{-1}~\mathrm{M}_\odot$ that have a stellar component, 
as well as the cosmic star formation rate density~(SFRD). In both of these cases, there are no significant deviations between the warmest and coldest subgroups that cannot be accounted for by cosmic variance or differences in the astrophysics and cosmological parameters.}

Clearly, the imprint of WDM mass on the halos resolved within the uniform-box simulations is rather weak.  
We note that the simulation-to-simulation scatter may be larger than current uncertainties on baryonic feedback due to the large range over which the parameters are varied.
However, the WDM signal at the mass scales probed in these simulations only imprints a $\sim 30\%$ reduction in the halo mass function when averaged over 100 simulations.
Since the WDM signal is so weak, these simple statistics on their own cannot be used to infer WDM particle masses.

\begin{figure*}
    \centering
    \includegraphics[width=\textwidth]{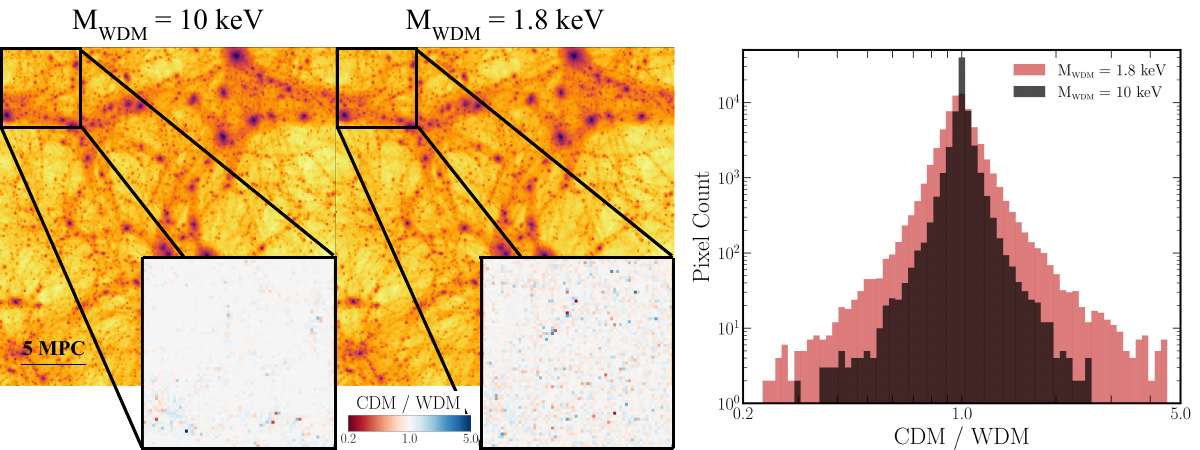}
    \caption{A comparison of the 2D projected total matter density for two different WDM models with M$_{\scriptscriptstyle{\rm WDM}} = 10$~(1.8)~keV in the left and middle panel, respectively. These particular images are not taken from the DREAMS suite as they only contain differences in the WDM particle mass and do not include variations in the initial density field, baryonic feedback, or cosmology.
    The size of each pixel covers a $10\times10~h^{-1}$~kpc area with a projected depth of 5~Mpc. 
    The top left corner of each image is enlarged and we provide the pixel-by-pixel ratio of each with the corresponding image from a CDM simulation.
    The zoomed-in panels have a diverging color map where blue~(red) corresponds to over~(under)-dense regions. The dynamic range is fixed with a maximum variation of $5\times$ to highlight the relative deviations from CDM. The right-most panel shows a histogram of the pixel ratios for the two WDM models.  In general, the 10~keV WDM model exhibits fewer pixel-level deviations from CDM than the 1.8~keV one. 
    } 
    \label{fig:im_comparison}
\end{figure*}

However, more information about the WDM is present in these simulations than might otherwise be captured by the statistics used in e.g., Fig.~\ref{fig:box_stats}, and advanced analysis methods may be more successful at extracting the signal.
To illustrate this point, consider how much WDM ``signal'' is present in a given area within the simulation.
Figure~\ref{fig:im_comparison} compares two simulations with the same initial density field, cosmology, and astrophysics, but different WDM mass (1.8 versus 10~keV).\footnote{These two examples are not taken from the DREAMS suite and are provided to illustrate differences when the initial conditions and baryonic/cosmological parameters are fixed.}  
The left and middle panels show the 2D projection of the matter density field from each simulation.
For each image, a section from the upper left corner is expanded to show the ratio of this image with the same one taken from a CDM simulation.
The ratio is displayed on a pixel-by-pixel basis with a diverging color map where less~(more) dense regions in the WDM projections relative to CDM are colored red~(blue).
Each pixel represents a $10 \times 10~h^{-2}~\mathrm{kpc^2}$ area with a depth of 5~$h^{-1}$~Mpc projected into two dimensions.
The dynamic range of the color map is matched for both the M$_{\scriptscriptstyle{\rm WDM}} = 1.8$ and 10~keV case such that the strongest colors align with a 5$\times$ deviation in either suppression or elevation.
The right-most panel of Fig.~\ref{fig:im_comparison} shows a histogram of the pixel ratios from the entire image for each WDM simulation.

Overall, there is less deviation from CDM for the  M$_{\scriptscriptstyle{\rm WDM}}=10$~keV model compared to the 1.8~keV case.
For the 10~(1.8)~keV image, 90\%~(66\%) of the pixels are consistent with CDM to within 10\%.  
Additionally, significant pixel deviations from CDM are more widespread for the 1.8~keV image compared to the 10~keV case, where they are localized around high-density regions.  These more subtle differences between images can potentially be captured by a CNN, as described next.  

\subsection{Inference with Convolutional Neural Networks}
\label{sec:CNN_results}

CNNs offer a natural way to predict DM properties from simulation data since they are designed to process 2D images within an inference framework.
The ability of these networks to move from simulated to observed data is currently limited and actively researched \citep[for a review, see][]{2019Baron}.
However, these models have been shown to successfully estimate the mass of clusters \citep{2019Ntampaka, 2023Ho}, distinguish between modified gravity and $\Lambda$CDM cosmologies \citep{2019Peel}, infer cosmological parameters from weak-lensing data \citep{2018Gupta, 2019Ribli, 2019Fluri}, create a mapping from baryon to DM  distributions \citep{2019Zhang}, and to locate substructure and estimate model parameters within strong lensing~\citep{2020Alexander, 2020Diaz, 2021Schuldt}.

We adopt the architecture and training procedures from the CAMELS Multifield Dataset~(CMD)~\citep{CMD} to train CNNs on field-level astronomical data.
This architecture was used in~\cite{2021VillaescusaA, 2021Villaescusa} and found to perform well at predicting cosmological parameters from astrophysical images.
In \cite{2024Rose}, this architecture was used to predict the WDM particle mass from similar images to those used here, except taken from N-body simulations where the WDM particle mass, $\Omega_{\rm m}$, and $\sigma_8$ were varied simultaneously.
We note that the images we analyze are created from full field-level data so there are no selection criteria imposed on the data.
Hence, this dataset includes low-mass halos, but we do not expect unresolved features of these halos to affect our results given the pixel resolution of the images,
Appendix~\ref{app:CNN} provides a brief summary of the key features of the architecture and image creation; the interested reader can go to~\cite{CMD} for a more detailed description.

\begin{figure}
    \centering
    \includegraphics[width=\columnwidth]{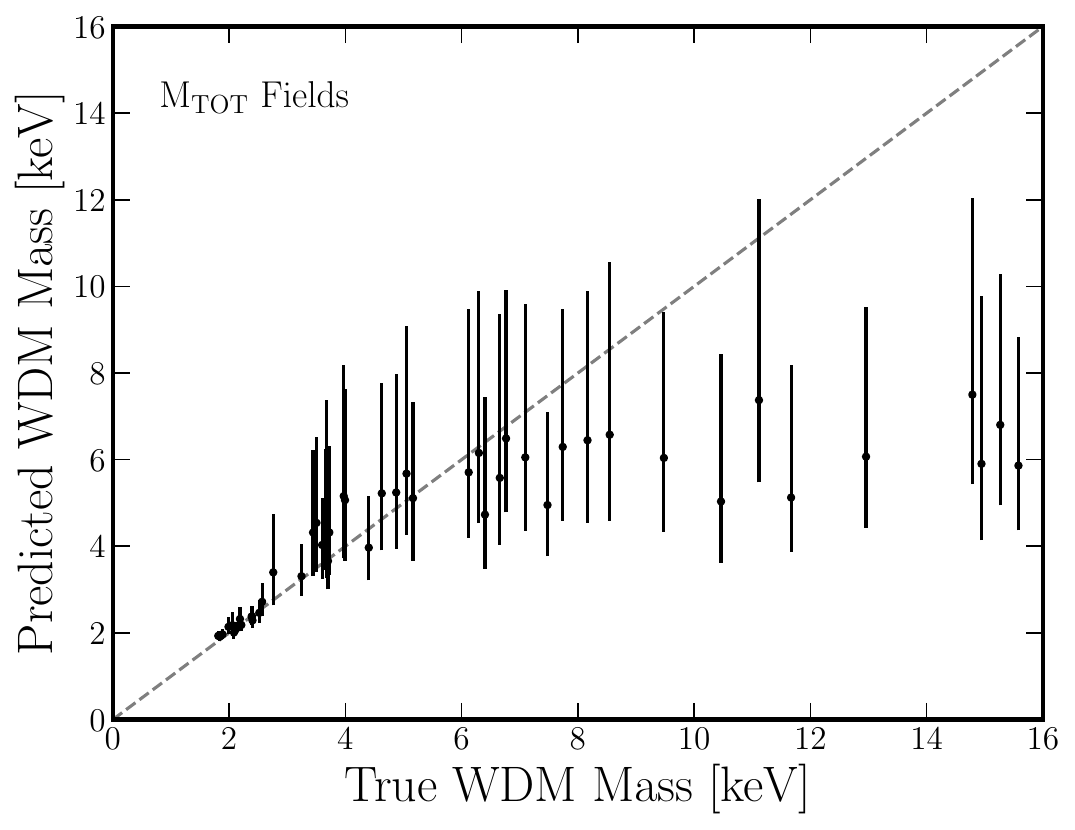}
    \includegraphics[width=\columnwidth]{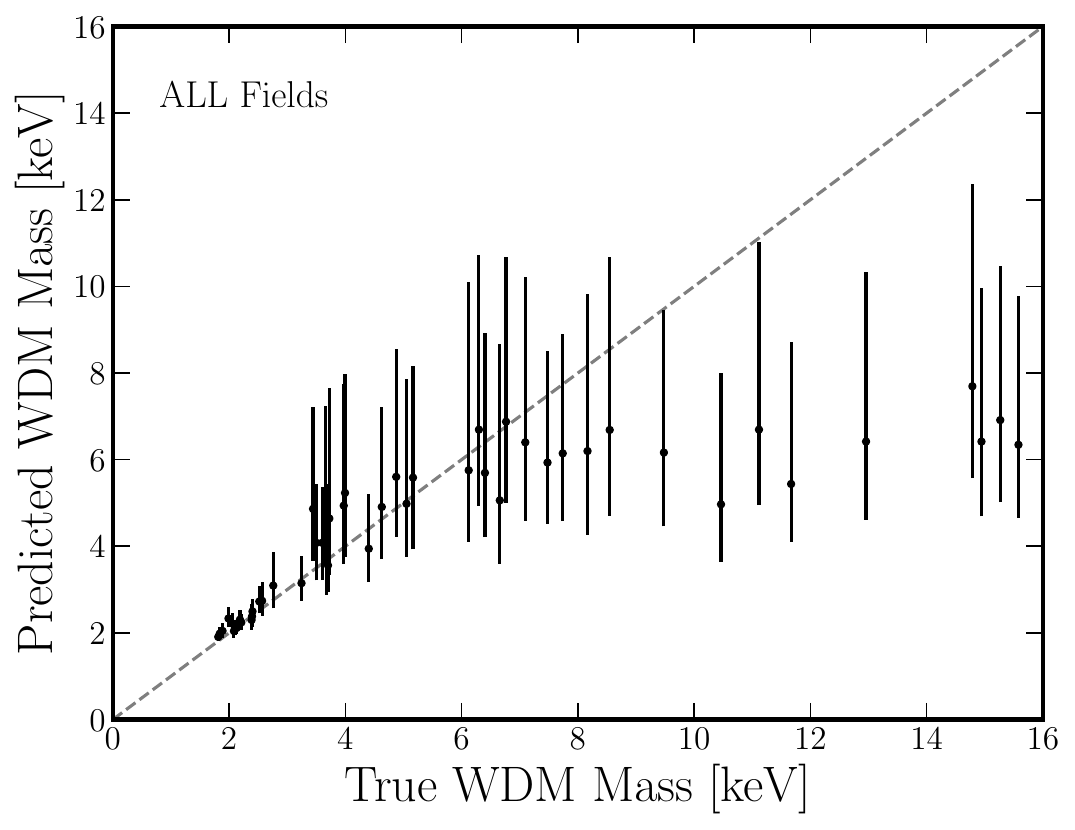}
    \includegraphics[width=\columnwidth]{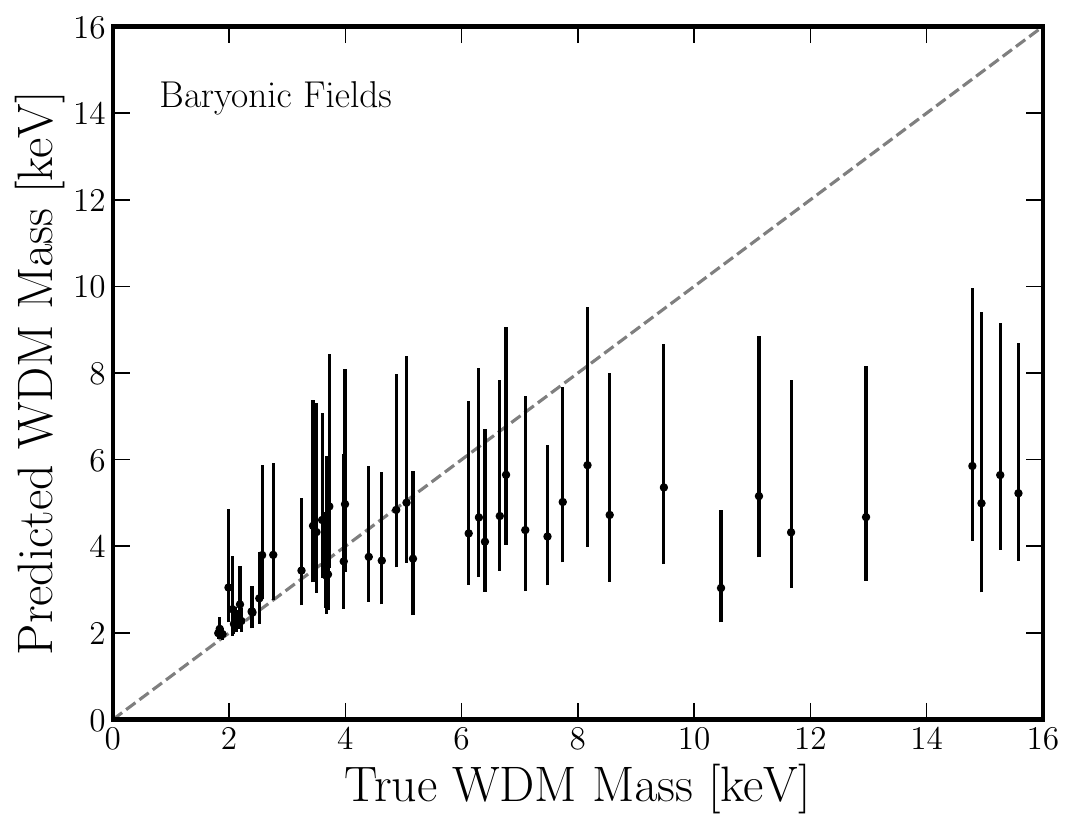}
    \caption{The results of the CNN trained to infer the WDM particle mass from fields available in the WDM uniform-box simulation, such as the total mass density~(top) and the baryonic fields~(bottom).  The middle panel shows the results using all the fields listed in Fig.~\ref{fig:box_images}. 
    Including the baryonic fields does not provide better predictive power over just the total matter density field.
    }
    \label{fig:wdm_inference}
\end{figure}

We first train and test the CNN to predict the WDM particle mass from images of the total matter density field.
The results of this analysis are shown in the top panel of Fig.~\ref{fig:wdm_inference}.
Each point on the plot shows the predicted mean of the posterior distribution from one image from the test dataset.
The errorbars show the $1\sigma$ uncertainty, which includes both the standard deviation of the posterior distribution~(aleatoric uncertainty) and the uncertainty intrinsic to the machine learning method~(epistemic uncertainty).
The epistemic uncertainty is always subdominant to the aleatoric uncertainty.
Since the simulations are sampled uniformly in $\mathrm{M_{\scriptscriptstyle{\rm WDM}}^{-1}}$, the posterior inferred by the CNN also has units of keV$^{-1}$.
To translate the uncertainties from units of keV$^{-1}$ to keV, we sample the uncertainty---the sum of the aleatoric and epistemic errors---as a Gaussian distribution with 1000 samples.
We then take the inverse of these samples and present the $1\sigma$ spread.

The trained neural network can predict the WDM particle mass up to 3.25~keV with less than $\pm 1$~keV uncertainty.
At greater DM particle masses, the network becomes more uncertain with some predictions being close to the true values and others plateauing at $\sim~6$~keV.
The plateau at $\sim~6$~keV results from the network being unsure of the true particle mass above $\sim 4$ so the network predicts the mean of the remaining distribution. 
This first analysis is restricted to only the total matter density field to mirror that of~\cite{2024Rose}. 
They found that their neural network (which utilized the same architecture used here) can make predictions with at least 2$\sigma$ confidence up to 3.5~keV with an average uncertainty of 0.6~keV.
Over the same mass range, our model has a larger average uncertainty of 1.2~keV.
Hence, the inclusion of a galaxy formation model (with parameter variation) does result in an increased uncertainty on the predicted WDM particle mass.\footnote{The analysis done in \cite{2024Rose} was performed on images with a resolution of 512$^2$ pixels whereas this study is done on images with a 256$^2$ pixel resolution.
Given that \cite{2024Rose} showed that pixel resolution did not have a significant effect on their results for their fixed-cosmology dataset at the simulation resolution used for our simulations, this should not have a significant effect on our results.}

We can attempt to infer the WDM particle mass from more than just the total mass density field given that the IllustrisTNG model is used in the simulations.
To test this, we train a CNN on a multifield dataset where each property in Fig.~\ref{fig:box_images} is included as a separate channel.
The results are provided in the middle panel of Fig.~\ref{fig:wdm_inference}. The trained CNN  predicts the WDM particle mass from the multifield dataset up to 3.25~keV with an uncertainty of less than 1~keV, matching the results when the model was trained on only the total matter density field.
Hence, including the baryonic and additional DM information does not significantly improve the ability to recover the WDM mass. This result suggests that any information that is contained in the baryonic fields is degenerate with the information in the total matter density field.

To test this further, we also train a model on the baryonic fields alone.
This includes all fields from Fig.~\ref{fig:box_images}, except $\mathrm{M_{TOT}}$, $\mathrm{M_{DM}}$, and $\mathrm{V_{DM}}$.
These results are shown in the bottom panel of Fig.~\ref{fig:wdm_inference}. 
In this case, the network predicts the WDM particle mass up to 2.0~keV with an uncertainty of less than 1~keV---slightly worse than the previous two examples. Therefore, while the baryonic fields do contain some information that can be used to predict the WDM particle mass, the total matter density field provides more.

To investigate which of the fields in Fig.~\ref{fig:box_images} contributes the most to the predictions in Fig.~\ref{fig:wdm_inference}, we train separate CNNs on each and compare the relative error of the results for WDM masses below 6.0~keV. 
We define the relative error as the difference between true and predicted values divided by the true value averaged over all points in the given dataset.
The fields $\mathrm{M_{TOT}}$ and $\mathrm{M_{DM}}$ produce comparable results with a relative error of 0.10, the best of all the cases tested here.
The next best results come from gas properties that all produce similar results, similar to those shown in the bottom panel of Fig~\ref{fig:wdm_inference}.
These fields (ne, HI, T, P, and $\mathrm{M_{GAS}}$) have a relative error of 0.15.
The rest of the fields ($\mathrm{V_{DM}}$, $\mathrm{M_{STAR}}$, Z, and B) all provide no predictive power with an average relative error of 0.95.

Overall, the most predictive power is contained in the DM density fields, which is degenerate with the information in the total matter density field.
Limiting the analysis to baryonic properties does still allow one to predict the WDM particle mass, although not as well. 
Nevertheless, the fact that the baryonic fields do have some discriminatory power is important as they are directly linked to experimental probes. For instance, the HI field is mapped by the 21 cm line~\citep{Pritchard_2012} and Lyman-alpha forest~\citep{Ouchi:2020zce}, the temperature field by X-rays~\citep{Schaye:1999vr, Haardt:2011xv} and the thermal Sunyaev-Zeldovich effect~\citep{Mroczkowski:2018nrv, Birkinshaw:1998qp}, and the B-field by Faraday rotation measures of radio sources~\citep{Ryu:2008hi, Feretti:2012vk}. To definitively say which of these probes is most useful in discriminating WDM masses, one would need to translate the simulation output into mock observational data, which is left for future work.

The results presented here represent the best predictions that can be made from the data supplied to the CNN, ie. 2D density fields taken from a $25 \times 25 \times 5$~($h^{-3}~\mathrm{Mpc^{3}}$) volume with a DM mass resolution of $7.81 \times (\Omega_{\rm m}/0.302) \times 10^{7}~h^{-1}~\mathrm{M}_\odot$.
Understanding how predictions from more observationally-motivated summary statistics, such as the power spectrum, compare to these predictions can shed light on the extent to which a given summary statistic, or ensemble, can capture all the information present in this dataset.
As shown in \cite{2024Rose}, the accuracy and uncertainty of these predictions improve with increased simulation resolution.
Therefore, the results that we show here represent a first step in the predictive power of this method where increasing the data size and resolution to current observational levels can help to understand whether our current observables miss important constraining information on the WDM particle mass.

\vspace{.5cm}

\section{Extensions to the DREAMS Project}
\label{sec:discussion}

We have presented two case studies that utilize data from \name to study properties of WDM while simultaneously accounting for astrophysical uncertainties.  
The framework illustrated through these examples consists of three modular components that can be adapted for a given problem of interest.  These include the choice of~(1) DM model, (2)~galaxy formation model, and (3)~astrophysical target.  This section discusses how these separate components can be altered in future DREAMS suites.

\subsection{Varying the Dark Matter Model}
\label{sec:DM_models}

Despite decades of effort, very little is known about the particle-physics nature of DM, including how it interacts with the Standard Model or with itself. DM may be fermionic or bosonic; it can behave as a wave or as a particle; it may interact with the Standard Model via new dark forces, or only via gravity; it may be one single species, or populate a rich ``dark sector”; it may be produced thermally from a hot plasma, or via non-thermal mechanisms. The possibilities are numerous and the objective of the DREAMS project is to take a strategic approach, focusing on several ``simplified models” that capture key features of the different classes of theories. The goal is to build intuition for how different DM properties impact galaxy formation and evolution, possibly unveiling new key observables.

The models we plan to explore as part of \name can be divided into two broad classes: those where the initial DM transfer function is modified and those with \emph{in situ} effects. The example discussed in this paper falls in the first category, as WDM is the simplest model that suppresses the DM transfer function. As a natural extension, we will consider more general modifications of the transfer function within the framework of the Effective Theory of Structure Formation~\citep[ETHOS;][]{2016Cyr-Racine,2016Vogelsberger}. Within the standard ETHOS example, the dark sector consists of a DM particle and a relativistic species~\citep[e.g., dark radiation,][]{Boehm:2001hm}.  The presence of dark radiation leads to dark-acoustic oscillations as well as DM self-interactions \citep{vandenAarssen:2012vpm}.  A useful variant on this framework is the $\hpkp$ phenomenological model proposed by~\cite{2020Bohr}, which isolates the early-universe ETHOS behavior by parameterizing modifications to the transfer function alone.

Separately, we will also consider simplified models that focus on the \emph{in situ} behavior of the DM. In this direction, one important question concerns the spin-statistics of DM: is DM a fermion or a boson? The DREAMS team will explore the possibility of simulating Wave Dark Matter~(i.e., ultralight bosons),  with a particular focus on Fuzzy Dark Matter (FDM)~\citep{Li:2018kyk, Dome:2022eaw, Lague:2023wes, Lora:2011yc, Mocz:2017wlg, Lancaster:2019mde, Dalal:2022rmp}, with or without self-interactions~\citep{Painter:2024rnc}. Particular care will be dedicated to studying the impact of baryonic feedback for FDM solvers, building on~\cite{Mocz:2019pyf}, which studied for the first time the interplay between baryons and FDM using \textsc{arepo}~\citep{Springel2010, Springel2019, Weinberger2020}.

As particle interactions in the Standard Model are rich and diverse, it is also plausible that new interactions arise within the dark sector. Another milestone of the DREAMS project will be the study of more complicated interactions between DM particles.  This includes elastic self interactions~\citep[SIDM;][]{SIDM-1999}, which have been routinely advocated as a solution to many of the small-scale ``tensions" mentioned in the introduction~\citep{2018PhR...730....1T}.  Even in relatively simple models with a single DM and mediator species, self-interactions of relevant strengths for galactic-scale observables can be achieved for sub-GeV DM and mediator masses~\citep{2016Kaplinghat}.  In general, these self interactions must be velocity-dependent to be consistent with current constraints from galaxy groups/clusters and MW satellites~\citep{2021arXiv210803243S, 2023MNRAS.518.2418S}.  Importantly, in the region of allowed parameter space, gravothermal collapse may play a key role in SIDM halo evolution~\citep{2002PhRvL..88j1301B, 2011MNRAS.415.1125K, 2019PhRvL.123l1102E, 2020PhRvD.101f3009N}, affecting galaxy diversity.  We will explore running hydrodynamical cosmological simulations of velocity-dependent SIDM, carefully considering numerical convergence of the halo's evolution under gravothermal collapse~\citep{2022Zeng,2024arXiv240201604M, Palubski:2024ibb}. 

Inelastic self interactions are also important to consider.  Both exothermic and endothermic DM scattering processes have been considered in the literature~\citep{Todoroki:2017pdh, 2019Vogelsberger, 2019MNRAS.483.4004T,2cDM-2022, 2021MNRAS.506.4421S, Shen:2022opd, 2023MNRAS.524..288O} and shown to affect, e.g.,  the central densities of galaxies.  Strongly dissipative dark sectors can have even more pronounced effects, leading to unexpected structures such as dark disks and/or novel compact objects~\citep{Fan:2013yva, 2018PhRvD..97l3018G, 2019JCAP...03..036C, 2022ApJ...934..122R}. Atomic Dark Matter~\citep{Kaplan:2009de,Cyr-Racine:2012tfp} provides an appealing simplified model for this case study, as it can also explain other conundrums, such as the DM-baryon coincidence~\citep{Bodas:2024idn}. Very recent works, which we plan to refine and further extend, have shown that even a small fraction of strongly dissipative DM can dramatically affect a galaxy's baryonic distribution~\citep{Roy:2023zar, Gemmell:2023trd}.

The simplified models outlined above illustrate promising avenues of exploration, which may be modified moving forward depending on computational feasibility (e.g. implementation strategy and/or  compute time). However, even if the specifics change, the guiding principle will be to work with a set of models that adequately capture the allowed range of DM phenomenology on galactic and sub-galactic scales. 

\subsection{Varying the Galaxy Formation Physics}
\label{sec:codes}

This work introduces two WDM suites with {\sc arepo} that utilize the TNG model.
Within this framework, we vary two parameters related to SN feedback and one related to AGN feedback.
These two suites, which comprise 2048 simulations, comprise the largest number of alternative DM simulations that account for uncertainties in astrophysical parameters.
However, they are limited by the (small) number of astrophysical parameters varied over, as well as the TNG physics implementation.
We therefore identify two axes in which the DREAMS program can be extended: (i)~expanded variations within a given subgrid model and (ii)~considerations of alternative sub-grid models.
This subsection details recent developments in the field along each axis and how they can be incorporated within the DREAMS project.

Recently, \cite{2023Ni} introduced a new suite of simulations that simultaneously vary 28 parameters within the TNG model, including the three considered in this work.
Some example variations are: the slope of the initial mass function~(IMF), the minimum stellar mass that produces a SN, and the minimum SN wind velocity.
Other parameters that they do not include in their analysis, such as the time of reionization, may mimic the effects of alternative DM models~\citep{2014Schultz, 2018LovellA, 2024Shen} or affect the formation of low-mass galaxies~\citep{2018Bose, 2018Dawoodbhoy, 2020Katz}.
While not all simulations within~\cite{2023Ni} match current observational constraints, the additional parameters were varied over ranges set to both be realistic and relevant for galaxy evolution scales. 
\cite{2023Ni} found that, while the additional variations resulted in a much larger range of galactic properties, their inference framework could still predict $\sigma_8$ and $\Omega_{\rm m}$ with some success.
Expanding the number of TNG parameters varied in the DREAMS suites would allow for the inclusion of e.g., uncertainties in the IMF slope or the start of reionization, which can affect small-scale observables.

However, even with these additional parameter variations, the simulations are still limited to the physics implemented within the TNG model.  As one example, TNG places a temperature floor on the interstellar medium~(ISM) that limits gas cooling below $10^4$~K.
This implementation does not allow for a multi-phase ISM, which can lead to more localized star formation and change how SN energy is injected into the system.
In lower-mass halos, such localized feedback can expel more mass from the center of the halo, reducing the central DM density~\citep{2010Governato, 2013Brooks, 2020MNRAS.497.2393L}.  This  coring from feedback 
has been proposed as a mechanism to alleviate some small-scale tensions, such as the diversity and too-big-to-fail problems~\citep{2014DiCintio, 2015Chan, 2016Tollet, 2020MNRAS.497.2393L}.
Introducing suites that are simulated with models that include processes like a multiphase ISM  allows one to bracket the uncertainties on small-scale structure predictions between different sub-grid implementations.  One such example that we plan to explore is the Adaptive Mesh Refinement~(AMR) code \textsc{ramses}~\citep{2002Teyssier}, which allows for localized stellar feedback modes.

Additionally, other codes take a different approach to solving the equations that govern galaxy formation and evolution.
\textsc{arepo} solves its equations of gravity using a TreePM grid~\citep{Bagla2002}.
While this method provides accurate results, \cite{2022Shao} showed that a Graph Neural Network~(GNN) trained on a halo catalog from one code does not always perform well when tested on a halo catalog from a different code.
For example, one GNN from~\cite{2022Shao} that is trained on a dataset from Gadget~\citep{2005Springel} does not predict $\sigma_8$ correctly when tested on a similar dataset from Enzo~\citep{2014Bryan}.
Similar to Arepo, Gadget also uses a TreePM grid to solve the equations of gravity.
Enzo on the other hand, is an AMR code that uses a fast Fourier technique to simulate gravity~\citep{1988Hockney}.
Other techniques used to solve gravity equations include the Fast Multipole Method~\citep{1987Greengard} implemented in PKDGrav3~\citep{2017Potter}, particle-mesh interactions implemented in CUBEP$^{3}$PM~\citep{2013Harnois}, and multipole-approximations implemented in Abacus~\citep{2021Garrison}.
It is unclear the extent to which the subtle numerical deviations between these simulations affect the field-level results inferred from them.

Since these deviations are present in gravity calculations, they should be more pronounced once the hydrodynamic equations are included as well.
The CAMELS project has tried to use machine learning to marginalize over numerical uncertainty from the galaxy formation implementation when predicting cosmological parameters~\citep{CAMELS}.
They do this by training a machine learning model on output from multiple simulation codes, then evaluate robustness by testing the model on output from yet another simulation code.
They concluded that adding simulation codes to the training dataset can provide more robust results than training on output from just one simulation code~\citep{2023deSanti, 2023Shao, 2023Ni}.
Because the DREAMS suites consist of many individual simulations packaged together, we can also expand the training datasets to include simulations from multiple codes to marginalize over numerical uncertainty.

\subsection{Varying the Zoom Target}
\label{sec:zoom_target}

Appendix~\ref{app:zoom-in} presents a procedure to efficiently generate large sets of zoom-in simulations, which was used to simulate the 1024 different MW-mass galaxies presented in Section~\ref{sec:emulation}.
Moving forward, we plan to also consider different targets for the zoom-in.  While MW-mass galaxies offer many opportunities to compare to local observations, dwarfs are DM-dominated systems that are particularly sensitive to changes in DM physics \citep[for a review, see][]{2019Simon}. 
Current and upcoming surveys, such as SPARC~\citep{2016Lelli}, DES~\citep{2005DES}, ELVES~\citep{2022Carlsten}, SAGA~\citep{2017Geha}, Roman~\citep{2015Spergel}, LSST~\citep{2009LSST}, ARRAKIHS \citep{2022Guzman}, and Merian~\citep{2023Luo}, offer a wealth of observations targeting dwarf systems that will benefit from theory investigations performed within the DREAMS project.

The local environment of the zoom-in galaxy is another consideration for future extensions of DREAMS. 
For the WDM zoom-in suite, there is an isolation criterion that removes systems with neighbors (within 1~Mpc) that are more massive than $5\times10^{11}~\mathrm{M}_\odot$. 
This requirement was put in place due to the extensive computational time (an increase of at least $2\times$) needed to simulate such systems.
To have more accurate MW analogs, we eventually want to simulate systems with a massive neighbor, similar to Andromeda.  
One promising technique is to augment a MW-mass zoom-in dataset with a smaller suite that focuses on these large-neighbor systems.  
Applying transfer learning methods, most of the training can be performed on the large and computationally inexpensive suite, with the machine-learned model then fine-tuned to incorporate results from the smaller and more expensive suite.

These augmented suites can also be used to extend the datasets to other more computationally intensive suites that would otherwise not be possible to execute for 1000+ simulations.
Another example would be a suite of simulations at higher resolution. This would allow one to investigate newly-resolved regimes at a reasonable computational cost, such as ultra-faint satellites~\citep{2015Bonnivard, 2016Brandt, 2020Hoof} or stellar streams~\citep{2021Banik, 2022Pearson}.

While the WDM zoom-in suite targets MW-mass haloes with a mass of $(8.8\pm1.2) \times 10^{11} h^{-1}~\mathrm{M}_\odot$, an additional suite of simulations could vary this target threshold.
The advantage of this would be to not rely on the MW's halo mass estimates and to incorporate DM measurement uncertainties to match current observations.
Additionally, if the mass range is extended further, the dataset could be used to compare to an ensemble of observations ranging from clusters to ultra-faint galaxies simultaneously.
This suite could help to incorporate a range of observables---from the density profiles of ultra-faint galaxies to strong-lensing predictions from clusters---self-consistently to thoroughly test various DM models.

\vspace{.5cm}

\section{Conclusions}
\label{sec:conclusion}

We introduced \name as a new approach to exploring the impact of DM on galaxy formation and evolution. Two WDM simulation suites---one consisting of uniform boxes and the other of MW zoom-ins---were introduced. Each suite includes 1024 cosmological hydrodynamic simulations generated with the \textsc{Arepo} code that assumes the TNG galaxy formation model. In addition to varying over the WDM particle mass, both suites also vary over the initial density field as well as several key parameters in SN and AGN feedback.  In this way, the simulations span uncertainties in both cosmic variance and astrophysics modeling.  The uniform-box suite also includes variations in cosmological parameters.  Notably, the DREAMS project includes the largest number of MW zoom-ins that have been generated to date, providing unprecedented statistics on the properties of classical satellites in a WDM cosmology.

We presented the results of two illustrative analyses that demonstrated the possibilities for machine-learning-focused studies. In summary:
\begin{itemize}[
  align=left,
  leftmargin=\parindent,
  itemindent=0pt,
  labelsep=0.5pt,
  labelwidth=1em,
  itemsep=0.5em]
    \item A neural network emulator was used to disentangle the multidimensional parameter space spanned by the simulations to quantify how the WDM mass and feedback parameters affect MW satellite counts. The analysis framework allows one to quantify how the spread in satellite counts from cosmic variance increases once uncertainties from astrophysical modeling are included.
    See Section \ref{sec:emulator_results}.
    \item CNNs were used to marginalize over uncertainties in cosmology and astrophysics to predict the WDM mass from two-dimensional density projections. While the DM and total matter density fields are the best predictors of the WDM particle mass, various gas properties including the electron number density, neutral hydrogen density, temperature, and pressure also contain predictive power. See Section~\ref{sec:CNN_results}.
\end{itemize}

Ultimately, \name will consist of thousands of cosmological hydrodynamic simulations that cover a range of systems from galaxy clusters to ultra-faint dwarfs. The long-term goals of the project are to better constrain the properties of DM, understand the dependence of galaxy formation on variations in the dark sector, and understand whether changes to baryonic feedback or dark sector physics can alleviate the small-scale tensions. There are three key extensions that are anticipated for future suites of the DREAMS project.  First, is the consideration of CDM and other alternative DM models.  Second, is the generalization of the galaxy formation model, both by including more free parameters within TNG and by also extending to other models beyond TNG. 
Third, is the inclusion of zoom-in simulation suites of different astrophysical targets.  
Taken together, these simulation suites will facilitate the comparison of DM predictions with current and upcoming data, enabling robust tests of CDM using small-scale galactic observables.

\section*{Acknowledgements}

The DREAMS team thanks the Simons Foundation for their support in hosting and organizing a foundational workshop on the early development of the DREAMS project.
JCR acknowledges support from the University of Florida Graduate School’s Graduate Research Fellowship and the Simons Foundation predoctoral program.
PT acknowledges support from NSF grant AST-2346977.
The work of FVN is supported by the Simons Foundation and by NSF grant AST 2108078. ML, SR, and KEK are supported by the Department of Energy~(DOE) under Award Number DE-SC0007968.  ML also acknowledges support from the Simons Investigator in Physics Award. MM acknowledges partial support by the National Science Foundation under Grant No. PHY-2010109. FYCR acknowledges the support of program HST-AR-17061.001-A whose support was provided by the National Aeronautical and Space Administration (NASA) through a grant from the Space Telescope Science Institute, which is operated by the Association of Universities for Research in Astronomy, Incorporated, under NASA contract NAS5-26555.
DAA acknowledges support by NSF grant AST-2108944, Simons Foundation Award CCA-1018464, and Cottrell Scholar Award CS-CSA-2023-028 by the Research Corporation for Science Advancement. KEK is supported by the NSF Graduate Research Fellowship Program under Grant No. DGE-2039656. Any opinions, findings, and conclusions or recommendations expressed in this material are those of the author(s) and do not necessarily reflect the views of the National Science Foundation.
Preliminary work that led to the creation of \name utilized Anvil at Purdue University through allocation PHY220051 from the Advanced Cyberinfrastructure Coordination Ecosystem: Services \& Support (ACCESS) program, which is supported by National Science Foundation grants \#2138259, \#2138286, \#2138307, \#2137603, and \#2138296.

\section*{Data Availability}

The data and code used to produce this paper can be made available upon reasonable request to the corresponding author.

\clearpage

\bibliography{citations}
\bibliographystyle{aasjournal}

\appendix

\section{Zoom-In Procedure}
\label{app:zoom-in}

\setcounter{equation}{0}
\setcounter{figure}{0} 
\setcounter{table}{0}
\renewcommand{\theequation}{A\arabic{equation}}
\renewcommand{\thefigure}{A\arabic{figure}}
\renewcommand{\thetable}{A\arabic{table}}

This work presents a suite of Milky Way~(MW) zoom-in simulations.  Creating this set of simulations is a multi-step process. 
In brief, the procedure is to:         
    (1)~simulate a uniform box with no baryons at a low resolution,
    (2)~perform a zoom-in simulation with no baryons at an intermediate resolution,
    and (3)~then perform a final zoom-in with baryons at the fiducial resolution.
This Appendix provides greater detail for each of these steps.

For the first step, the uniform-box simulation has a $(100~h^{-1}~{\rm Mpc})^3$ volume with $256^3$ particles.
This provides a mass resolution of $5 \times 10^9~h^{-1} \mathrm{M}_\odot$ and a spatial resolution of $4.9~h^{-1} {\rm kpc}$ at $z=0$.
At this resolution, a MW-mass halo consists of at least 300 particles, which is sufficient to locate the halo in the box, although not good enough to resolve its internal structure.  This simulation takes 200~cpu-hours to complete.

\newcommand{\widefootnote}[1]{
    \footnote{\rlap{\parbox{\dimexpr\textwidth-14pt\relax}{#1}}}
}

For the second step, we choose a random halo from this uniform box whose virial mass falls within $(1.58$--$1.61) \times 10^{12}~\mathrm{M}_\odot$. This range is toward the upper end of the observed MW mass estimates compiled in \cite{2016Bland}.\widefootnote{Due to the low resolution of the uniform box, the mass criterion is not maintained at higher resolution. The average mass of the final MW targets is  $(1.3 \pm 0.2) \times 10 ^{12}~\mathrm{M}_\odot$, still within the acceptable observational range~\citep{1999Wilkinson, 2010Watkins, 2014Kafle, 2018Sohn, 2019Eadie}.
In the future, additional care should be taken to choose halos that fall within the desired mass range by checking the isolation criterion after the intermediate N-body zoom-in simulation.}
We also apply an isolation cut, requiring that there are no other halos more massive than $7.2 \times 10^{11}~h^{-1} \mathrm{M}_\odot$ within $1~h^{-1} {\rm Mpc}$ of the selected halo.
This criterion is implemented for computational reasons as halos that are close to more massive halos take too long to simulate or have significant contamination from low-resolution particles within the halo at $z=0$.
Overall, this isolation criterion excludes $\sim 12\%$ of all MW-mass galaxies found in the uniform box. Future work will explore the possibility of removing this isolation criterion so that MW-analogs with large neighbors (like Andromeda) can be studied.

For the target halo, we select all particles within 5$\times$ its virial radius, trace their positions back to $z=127$, and define an ellipse that encompasses them.
This ellipse is then used to determine the high-resolution region of the intermediate N-body zoom-in simulation.
In this region, the mass resolution is $1.0\times 10^7~ h^{-1} \mathrm{M}_\odot$ and the spatial resolution is $0.61~h^{-1}\rm{kpc}$ at $z=0$.
The target halo contains $\sim 1\times 10^5$ particles, which allows for a better definition of its internal structure.
The low-resolution region fills the rest of the box.
Far from the target halo ($\gtrsim 10~{\rm{Mpc}}$ at $z=127$), the simulation has a mass resolution of $3~\times~10^{10}~h^{-1}~\mathrm{M}_\odot$.
Surrounding the buffer region, the simulation has concentric regions of decreasing mass that step from the low-resolution region to the high-resolution region.
On average, each N-body zoom-in simulation takes 250~cpu-hours, but can take over 1000~cpu-hours if the MW-mass halo is in a crowded environment.  As it turns out, this process is more efficient than simulating a uniform box at sufficient resolution to properly identify the internal structure of a target MW-mass halo.  As the zoom-in procedure requires over 1000 of these simulations, the increased computation time would be comparable to that spent on the final hydrodynamic suite.

We use the results from the N-body zoom-in simulation from the second step to create the initial conditions for the final hydrodynamic zoom-in simulation.  
In this process, there can be significant contamination from low-resolution particles in the intermediate N-body zoom-in due to the low resolution of the initial uniform box.
To produce a less-contaminated high-resolution region for the final zoom-in simulation, we ensure that we include all particles, high and low resolution, within $5 \times \mathrm{M}_\mathrm{vir}$ during this step.
We then calculate the ellipse that encloses all particles at $z=127$ and use this new ellipse to define the high-resolution region.

For $\sim$15\% of the simulations, this procedure produces initial conditions that can take an order of magnitude longer to simulate than average, due to a large high-resolution region.
For these simulations, which we define as having more than $10^7$ high-resolution particles,
we use a convex hull to define the region that encompasses the high-resolution particles.
The convex hull allows for a smaller high-resolution region while still encompassing all particles that will be in the MW-mass halo at $z=0$.

\cite{2014Onorbe} has shown that initial conditions created in convex hulls can result in greater contamination from low-resolution particles than ellipses.
Hence, we only use convex hulls when necessary for computational purposes.
We do find that some simulations that use the convex hull either result in significant contamination from low-resolution particles ($>$1\% by mass within the virial radius)
or still maintain a significant computational cost.  For these simulations, we implement a new method to reduce the high-resolution Lagrangian region and better target the area around the MW-mass halo.

The method begins by first tracing all particles within the halo's virial radius back to $z=127$.
Next, we calculate the distance from each of those particles to the center-of-mass of the high-resolution Lagrangian region at $z=127$.
For each particle, we also calculate the distance to its 1024$^\mathrm{th}$ neighbor at $z=127$,
only considering the particles that are within the virial radius at $z=0$.
We then remove any particle whose distance to the center of the Lagrangian region added to the distance to its $1024^\mathrm{th}$ neighbor is greater than 5$\times$ the average distance to the center of the Lagrangian region. This cut removes any low-resolution particles that end up within the virial radius of the target halo, but that started from a much further distance. The procedure does not remove most low-resolution particles that are pulled from nearby regions.  

After removing the outlier particles, we rebuild the initial conditions around the MW-mass galaxy to maintain a reasonable buffer between the high-resolution and low-resolution regions. 
To do this, we first trace all particles within 5$\times$ of the target halo's virial radius (the same particles used previously to define the target halo) back to $z=127$. 
For every particle that passes the cut described above, we find the 1024$^\mathrm{th}$ closest particles from those within $5 \times$ the virial radius.
Including both the particles that pass the cut and their selected neighbors, we redraw an enclosing ellipse around these particles to make the ICs for the hydrodynamic simulation.

\vspace{0.2in}
\section{Warm Dark Matter Implementation}
\label{app:DM}

\setcounter{equation}{0}
\setcounter{figure}{0} 
\setcounter{table}{0}
\renewcommand{\theequation}{B\arabic{equation}}
\renewcommand{\thefigure}{B\arabic{figure}}
\renewcommand{\thetable}{B\arabic{table}}

\begin{figure}
    \centering
    \includegraphics[width=0.5\columnwidth]{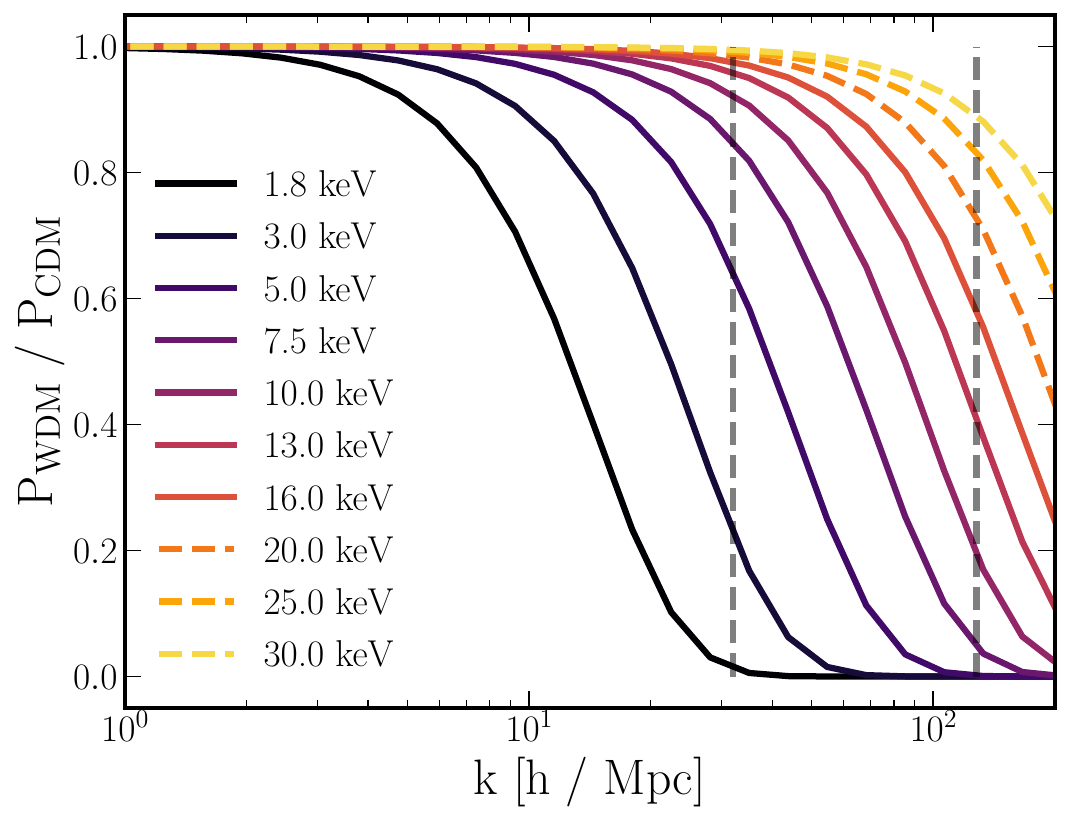}
    \caption{The suppression of power relative to CDM for the WDM particle masses used in the DREAMS suites, plotted as a function of comoving wavenumber. The vertical grey-dashed line indicates the Nyquist frequency for the uniform~(zoom) simulations at $32~(128)~h~ {\rm Mpc}^{-1}$.
    The colored dashed lines ($>16~{\rm keV}$) are only included in the zoom-in suite as they are indistinguishable from CDM in the uniform-box suite.}
    \label{fig:wdm_transfer}
\end{figure}

Figure~\ref{fig:wdm_transfer} illustrates the suppression of the Warm Dark Matter~(WDM) power spectrum relative to Cold Dark Matter~(CDM), as a function of comoving wavenumber. 
Each colored line corresponds to a different WDM mass, starting from 1.8~keV and extending to 30~keV. 
The vertical dashed gray lines indicate the Nyquist frequencies for the uniform-box~($32~h~{\rm Mpc}^{-1}$) and zoom-in~($128~h~{\rm Mpc}^{-1}$) WDM simulations---see Tab.~\ref{tab:simulations} for further details. 
At the uniform-box resolution, WDM masses greater than 16~keV are indistinguishable from CDM and are thus not included in this suite. 
The zoom-in suite includes masses up to 30~keV, as indicated by the dashed colored lines in Fig.~\ref{fig:wdm_transfer}, to ensure that inference performed on this suite includes a CDM-like prediction.
Note that WDM models with masses $\lesssim 2$~keV are strongly ruled out by multiple analyses \citep{2017Irsic, 2020Hsueh,2021Enzi, 2021NadlerA, 2021Nadler}, but are included in the suites to help machine learning models learn the characteristic features that distinguish WDM from CDM. 

To simulate thermal-relic WDM, we apply a suppression to the initial matter power spectrum compared to CDM. 
The relationship between the WDM power spectrum, $P_{\rm WDM}(k)$, and CDM power spectrum, $P_{\rm CDM}(k)$, depends on the wavenumber, $k$.  Following \cite{2001Bode}, we take $P_{\rm WDM}(k)=\beta(k) P_{\rm CDM}(k)$ with 
\begin{align}
\label{eq:wdm}
    \beta(k) &= \left( \left( 1 + (\alpha k)^{2.4} \right)^{-5.0/1.2} \right)^2 \quad \text{and}\\
    \alpha &= 0.048~h^{-1}~\rm{Mpc} \left(\frac{M_{\scriptscriptstyle{\rm WDM}}}{\mathrm{keV}}\right)^{-1.15} \left( \frac{\Omega_{\rm m} - \Omega_{\rm b}}{0.4} \right)^{0.15} \left(\frac{h}{0.65} \right)^{1.3}   \, .
\end{align}
Here, $\mathrm{M_{\scriptscriptstyle{\rm WDM}}}$ is the WDM particle mass in keV, 
$\Omega_{\rm m}$ is the total matter density, and $\Omega_{\rm b}$ is the baryon density.
The value of $\Omega_{\rm m}$ is fixed to $0.302$ for the zoom-in suite and 
is varied according to a Sobol sequence within the range $\Omega_{\rm m} \in [0.1,0.5]$ for the uniform-box suite.

Some simulations of WDM include additional velocity kicks to the simulation particles at early times \citep[e.g.,][]{2007Wang, 2012Maccio, 2013Viel}.
These kicks are meant to approximate the additional thermal velocities of the DM particles.
\cite{2017Leo} investigated whether kicks accurately capture the underlying population they are trying to model by examining how including kicks and varying simulation resolution affect power at small scales.
They found that including velocity kicks in the simulations introduces greater numerical noise and artificial fragmentation than if they were excluded.
Therefore, we do not include thermal kicks in our simulations to reduce numerical noise.  

Artificial fragmentation can occur in WDM simulations due to artificial power at small scales present from discrete sampling.
During the evolution of the simulation, this artificial power can grow to the size of a small halo---a so-called spurious halo.
Using the equation for the halo masses expected from artificial fragmentation from \cite{2007Wang}, we estimate a maximum mass of $\sim 8 \times 10^8~h^{-1}~\mathrm{M}_\odot$ for the spurious halos in the WDM uniform-box simulation (assuming an extreme model with $\Omega_{\rm m} = 0.5$ and $\mathrm{M}_{\scriptscriptstyle{\rm WDM}} = 1.8$~keV). 
The mass resolution for the WDM uniform box with $\Omega_{\rm m} = 0.5$ is $1.3 \times 10^8~h^{-1}~\mathrm{M}_\odot$.
Therefore, the scale of spurious structure formation in this case is $\sim 6$ particles.
Because the image-making procedure smooths the density field over a kernel of 32 particles,  the artificial fragmentation should not significantly affect the results.

For the zoom-in simulations, the maximum mass for the spurious halos with 1.8 keV WDM is $\sim 7 \times 10^7~h^{-1}~ \mathrm{M}_\odot$.
These simulations have a mass resolution of $1.2 \times 10^6~h^{-1}~\mathrm{M}_\odot$, leading to spurious structures of $\sim 60$ particles.
These structures are larger, in total particle count, than those in the uniform-box simulation and may play a more pronounced role in the galaxy evolution for the most extreme WDM models.
The analysis in Section~\ref{sec:emulation} only includes galaxies with $\mathrm{M_{halo}} > 10^8~h^{-1}~\mathrm{M}_\odot$ and therefore should not be affected by the presence of artificial halos. 

\vspace{0.2in}
\section{Machine Learning Architectures}
\label{app:ML}

\setcounter{equation}{0}
\setcounter{figure}{0} 
\setcounter{table}{0}
\renewcommand{\theequation}{C\arabic{equation}}
\renewcommand{\thefigure}{C\arabic{figure}}
\renewcommand{\thetable}{C\arabic{table}}

This section provides details of the machine learning architectures used for the analyses presented in the main body of the paper.

\subsection{Emulator}
\label{app:emulator}

Section~\ref{sec:emulation} presents a first application of the WDM zoom-in suite, using emulation to demonstrate how satellite count is affected by WDM particle mass as well as the SN and AGN parameters varied in this study.  The emulator consists of multiple layers of fully-connected nodes.
The first layer comprises the inputs to the network:  WDM particle mass, $\epsilon_w$, $\kappa_w$, and $\epsilon_{f, {\rm high}}$. 
Recall that the cosmological parameters are kept fixed for the MW zoom-in suite.
The first layer is followed by a set of layers, each consisting of neurons that are fully connected to the subsequent layer.
The number of neurons in each layer, as well as the total number of layers, are tuned as hyperparameters.
The final layer of the network is the outputs of the neural network, which correspond to the mean and standard deviation of MW-mass satellite counts for the example in Section~\ref{sec:emulation}.

The emulator is trained on the MW-mass hosts in the suite, which are split into training, validation, and testing sets, each comprising 90\%, 5\%, and 5\% of the total, respectively.
We find that increasing the size of the training and testing sets does not significantly affect the results.
Training is performed over 500 epochs using the loss function \citep{CMD, moments_network}
\begin{equation}
\label{eq:loss}
    \mathcal{L} = \left( \sum_{i \epsilon \mathrm{batch}} \left(\theta_i - \mu_i\right)^2 \right) + \left( \sum_{i \epsilon \mathrm{batch}} \left(\left(\theta_i - \mu_i\right)^2 - \sigma_i^2 \right)^2 \right) \, ,
\end{equation}
where $\theta_i$ is the true simulation parameter of the $i^\mathrm{th}$ simulation in the batch used to train the model and 
$\mu_i$~($\sigma_i$) are the mean~(standard deviation) of the posterior given as model outputs during training. 
The number of epochs is tuned from 300 to 1000 epochs, over which there were no significant effects on the results.

The hyperparameters of the model are the number of fully connected layers~(NL), the number of nodes in each fully connected layer~(NN), the learning rate~(LR), the weight decay~(WD), and the dropout rate~(DR). 
The LR determines how quickly the neurons within the machine learning model can change during training, the WD regularizes the network by imposing a penalty for large neuron weights, and the DR is the fraction of connections between neurons that are not used in a given epoch during training.
These parameters are varied within these ranges:
\begin{eqnarray} \nonumber
    \mathrm{NL} &\in& [1, 5] \\ \nonumber
    \mathrm{NN} &\in& [4, 10^3] \\ \nonumber
    \mathrm{LR} &\in& [10^{-5}, 10^{-1}] \\ \nonumber
    \mathrm{WD} &\in& [10^{-8}, 1] \\ \nonumber
    \mathrm{DR} &\in& [0.2, 0.8] \, .\nonumber
\end{eqnarray}
The LR, WD, and DR are all sampled logarithmically during hyperparameter tuning.
A unique value of NN is chosen for each layer within the given network.
Hyperparameter tuning is done with \textsc{optuna} \citep{optuna} over 50 trials.

The standard deviation of the posterior provides the aleatolic error on the result.
To include the epistemic error, we average the results of an ensemble of trained models.
For the results presented in Section~\ref{sec:emulator_results}, we independently train 10 instances of the emulator with the same tuned hyperparameters but different initialization of weights.
The mean values presented correspond to the mean result across all ten instances of the emulator.
The aleatoric uncertainty is taken as the average width of the posterior from the ten instances of the network.
The epistemic uncertainty is the standard deviation of the mean of the posterior across all ten instances of the emulator.
The uncertainty that we report is the aleatoric uncertainty added in quadrature to the epistemic uncertainty.

For all results presented in Section~\ref{sec:emulator_results}, the aleatoric uncertainty is greater than the epistemic uncertainty.
On average, the aleatoric uncertainty is $\sim~7~\times$ greater than the epistemic uncertainty.
However, there is a trend with WDM particle mass where larger masses result in a larger epistemic uncertainty, but still less than the aleatoric uncertainty.

The right panel of Fig.~\ref{fig:emulator_WDM} compares the output of the emulator with the simulation data to validate the results.
The emulator data~(black) represents the expected number of satellites and the uncertainty from both cosmic variance and astrophysical variations.
The simulation data~(orange) represents the average number and standard deviation of satellites for all simulations binned into 0.5~keV bins.
We include astrophysics variations, instead of limiting them to the fiducial values, to have a consistent comparison with 
the entire simulation dataset.

At low and moderate WDM masses (M$_{\scriptscriptstyle{\rm WDM}} \lesssim~12$~keV), the emulator and simulation results agree well.
At high masses, the simulation data becomes more noisy and oscillates around the average emulator output due to the lower number of simulations with large WDM mass.
There are only 95 simulations with a WDM particle mass greater than 12~keV and only 32 greater than 20~keV.
While these results confirm the accuracy of the emulator, we suggest caution in overinterpreting differences between models with WDM particle masses greater than 12~keV.

\subsection{Convolutional Neural Network}
\label{app:CNN}

Section~\ref{sec:inference} presents a first application of the hydrodynamic WDM uniform-box suite, using a convolutional neural network~(CNN) to predict the WDM particle mass from astrophysical images. The neural network takes as input 2D projected images of DM and/or baryonic density fields (see Fig.~\ref{fig:box_images} for examples).  Using the \textsc{voxelize} code,\footnote{\url{https://github.com/leanderthiele/voxelize}} each image is created from the simulation output by first reconstructing the 3D density grid, with a resolution of ($10~h^{-1}$~kpc)$^3$, for a property of interest. The algorithm used to recreate the density field first computes the distance from each simulation particle to its 32 nearest neighbors.
The value of 32 affects the smoothness of the reconstructed density field but is commonly used during this kind of reconstruction \citep[eg.][]{1989Hernquist, 1992Monaghan}.
Each particle is then represented by a uniform sphere whose density (associated with the used quantity) is deposited into a regular grid \cite[for details, see][]{CMD}.

After generating the 3D density grid, we produce the 2D images by taking $5~h^{-1}~{\rm Mpc}$-thick slices of the voxelized grid and projecting them along the short axis. The thickness of $5~h^{-1}~{\rm Mpc}$ is chosen to match that used for the CAMELS project~\citep{CAMELS} and for \cite{2024Rose}; it allows for a balance between a larger dataset and more information contained within each image.
We produce 15 unique images per simulation by taking 5 slices along each of the 3 axes of the simulation box and then projecting along the short axis of each slice.
Each image is then augmented using four $90^\circ$~rotations and two reflections.
Thus, there are 120 images per simulation for a total of 122,880 images for a suite containing 1,024 simulations.
Before training, we normalize the dataset by taking the log of all pixels, normalizing each pixel to the average value of the entire dataset, and then setting the standard deviation across all pixels to one.
This normalization is not done on an image-by-image basis, so relative differences between different simulations will persist in the images created from those simulations.

We create images for eleven different astrophysical properties to use during training.
These properties are:
\begin{tasks}(2)
    \task[] \makebox[1cm][l]{$\mathrm{M_{TOT}}$} - the total mass density. 
    \task[] \makebox[1cm][l]{$\mathrm{M_{DM}}$} - the DM mass density.
    \task[] \makebox[1cm][l]{HI} - the neural hydrogen mass density.
    \task[] \makebox[1cm][l]{ne} - the electron number density.
    \task[] \makebox[1cm][l]{$\mathrm{M_{GAS}}$} - the total gas mass density.
    \task[] \makebox[1cm][l]{T} - gas temperature.
    \task[] \makebox[1cm][l]{P} - gas pressure.
    \task[] \makebox[1cm][l]{$\mathrm{V_{DM}}$} - DM velocity modulus (speed).
    \task[] \makebox[1cm][l]{$\mathrm{M_{STAR}}$} - stellar mass density.
    \task[] \makebox[1cm][l]{Z} - gas metallicity.
    \task[] \makebox[1cm][l]{B} - magnetic field modulus.
\end{tasks}
See Fig.~\ref{fig:box_images} for examples of each property.

The network architecture consists of 19 convolutional layers, each followed by a batch normalization and a leaky ReLU activation function.
The overall size of each image is reduced from $256 \times 256$ pixels to $128 \times H$ nodes, where $H$ is a hyperparameter (see below).
The resulting nodes are then passed through a fully connected layer, with dropout, reducing the stack to $64 \times H$ nodes.
These nodes are then passed through a leaky ReLU activation function, dropout is applied again, and the nodes are passed through another fully connected layer to produce the simulation outputs.

The training, validation, and testing sets are split at the simulation level so images from one simulation are not used in both training and validation/testing. This ensures that the models see no overlap in the parameter space between the training and testing image sets.
We reserve 5\% of the simulations~(6,144 images) for the testing and validation sets each, then use the rest for the training set.
We have verified that reserving larger fractions for the validation and testing sets does not produce any substantial changes to the results.
The results that we present in Section~\ref{sec:inference} come from the testing set to ensure that our neural network has never seen the data used to produce these results.

This architecture takes a single image as input and outputs two values: the mean and standard deviation of the posterior distribution for the desired property of the simulation.
The mean and standard deviation of the posterior are calculated through the loss function given in Equation~\ref{eq:loss}.
We utilize a batch size of 128 images but find that other values between 32 and 1024 also produce similar results.
Training is done over 200 epochs.

During training, we tune four hyperparameters over 50 trials for each model.
In addition to the number of hidden layers, H, see above, the other hyperparameters include the learning rate~(LR), dropout rate~(DR), and weight decay rate~(WD).  The LR controls the average amplitude of the changes that are made to the model parameters between epochs.  The DR controls what fraction of nodes in the fully connected layer are removed during training.  The WD rate controls how quickly the parameters in the fully connected layer decay during the training process. The four hyperparameters are varied within the following ranges:
\begin{eqnarray} \nonumber
    \mathrm{LR} &\in& [10^{-5}, 5 \times  10^{-3}] \\ \nonumber
    \mathrm{DR} &\in& [0, 0.9] \\ \nonumber
    \mathrm{H} &\in& [6, 12] \\  \nonumber
    \mathrm{WD} &\in& [10^{-8}, 10^{-1}] \nonumber
\end{eqnarray}

The trained CNN model used in this analysis has the set of hyperparameters that produce the lowest value of the loss function when evaluated on the validation set.
For each of our models, we find that the training, validation, and testing losses are similar, which suggests that they do not suffer from overtraining.

To incorporate the uncertainties intrinsic to the machine learning method, we train an ensemble of five CNNs to measure the epistemic uncertainty.
Each CNN is retrained using the optimal hyperparameters from our hyperparameter tuning but with different initial parameter weights.
We take the aleatoric uncertainty as the average width of the posterior and the epistemic uncertainty as the standard deviation of the mean of the posterior from the five networks divided by the square root of the number of trained CNNs.
We then present the total uncertainty as the aleatoric uncertainty added in quadrature to the epistemic uncertainty.

\label{lastpage}
\end{document}